\documentclass[superscriptaddress,showkeys,twocolumn,nofootinbib,longbibliography]{revtex4-1}

\usepackage{amsmath}
\usepackage{amssymb}
\usepackage{graphicx}
\usepackage{epsf}
\usepackage{slashed}
\usepackage{enumitem}
\usepackage[czech,english]{babel}
\usepackage[cp1250]{inputenc}
\usepackage{hyperref}
\usepackage{xcolor}
\usepackage{amscd}

\usepackage[only,llbracket,rrbracket,llparenthesis,rrparenthesis]{stmaryrd} 
\usepackage{accsupp} 

\newcommand{\refer}[1]{(\ref{#1})}
\newcommand{\diff}{\mathrm{d}}

\usepackage{bbm} 

\usepackage[nodayofweek]{datetime}
\newdateformat{mydate}{\twodigit{\THEDAY}{ }\shortmonthname[\THEMONTH], \THEYEAR}
\usepackage[normalem]{ulem}

\begin{document}

\title{Mechanization of scalar field theory in (1+1)-dimensions: BPS mech-kinks and their scattering}

\author{Filip Blaschke}
\email{filip.blaschke(at)physics.slu.cz}
\affiliation{
Research Centre for Theoretical Physics and Astrophysics, Institute of Physics, Silesian University in Opava, Bezru\v{c}ovo n\'am. 1150/13, 746~01 Opava, Czech Republic.
}

\author{Ond\v{r}ej Nicolas Karp\'{i}\v{s}ek}
\email{karponius(at)gmail.com}
\affiliation{
Institute of Physics, Silesian University in Opava, Bezru\v{c}ovo n\'am. 1150/13, 746~01 Opava, Czech Republic.
}

\author{Luk\'a\v{s} Rafaj}
\email{lukasrafaj(at)gmail.com}
\affiliation{
Institute of Physics, Silesian University in Opava, Bezru\v{c}ovo n\'am. 1150/13, 746~01 Opava, Czech Republic.
}

\begin{abstract}
We present an updated version of a general-purpose collective coordinate model that aims to fully map out the dynamics of a single scalar field in (1+1)-dimensions. This is achieved by a procedure that we call a `mechanization': we reduce the infinite number of degrees of freedom down to a finite and controllable number by chopping the field into flat segments connected via joints. In this paper, we introduce two new ingredients to our procedure. The first is a manifestly BPS mechanization in which BPS mech-kinks saturate the same bound on energy as their field-theoretical progenitors. The second is allowing the joints to `switch', leading to an extended concept of the effective Lagrangian, through which we describe direct collisions of mech-kinks and anti-kinks.
\end{abstract}

\keywords{BPS, kinks, collective coordinate model, mechanization}

\maketitle



\section{Introduction}
\label{sec:I}

Field theories in (1+1)-dimensions with disconnected vacua support topological solitons -- kinks -- that are stable, particle-like objects. Kinks (and their higher-dimensional relatives) are relevant in many areas of contemporary physics, including cosmology, condensed matter and particle physics \cite{Manton:2004tk, Vilenkin:2000jqa, Shnir:2018yzp}. 

The collisions of solitons have become a major avenue for theoretical exploration of the inner workings of non-linear field dynamics. Indeed, during collisions, the non-linearity is `switched on' only intermittently and with an intensity that can be tuned, among other parameters, by the initial velocities of the impactors. The holy grail of soliton dynamics would be the ability to predict -- given the initial state of solitons and the model at hand -- the outcome of any collision.

 Although the kink-anti-kink ($K\bar{K}$) scattering have been studied since the late 70-ties \cite{Sugiyama:1979mi, Campbell:1983xu, Moshir:1981ja, Anninos:1991un, Belova:1985fg} the true quantitative understanding of their main characteristics has been achieved only recently \cite{Manton:2020onl, Manton:2021ipk, Adam:2021gat, Dorey:2011yw, Adam:2022mmm, Weigel:2013kwa, Dorey:2023izf, Adam:2023kel, Adam:2023qgx} (see also references in \cite{2019arXiv190903128K}). 

A hallmark feature of $K\bar{K}$ collisions is the bouncing phenomenon. It has been long since understood as a resonant transfer of kinetic energy to and from colliding solitons into localized modes of the field.
In the case of $\phi^4$ kink, they are the shape modes residing on the kinks themselves \cite{Manton:2021ipk}, while for $\phi^6$ model, a delocalized mode emerges in between the  $\bar{K}K$ pair  \cite{Adam:2022mmm}.

\begin{figure}[htb!]
\begin{center}
\includegraphics[width=0.99\columnwidth]{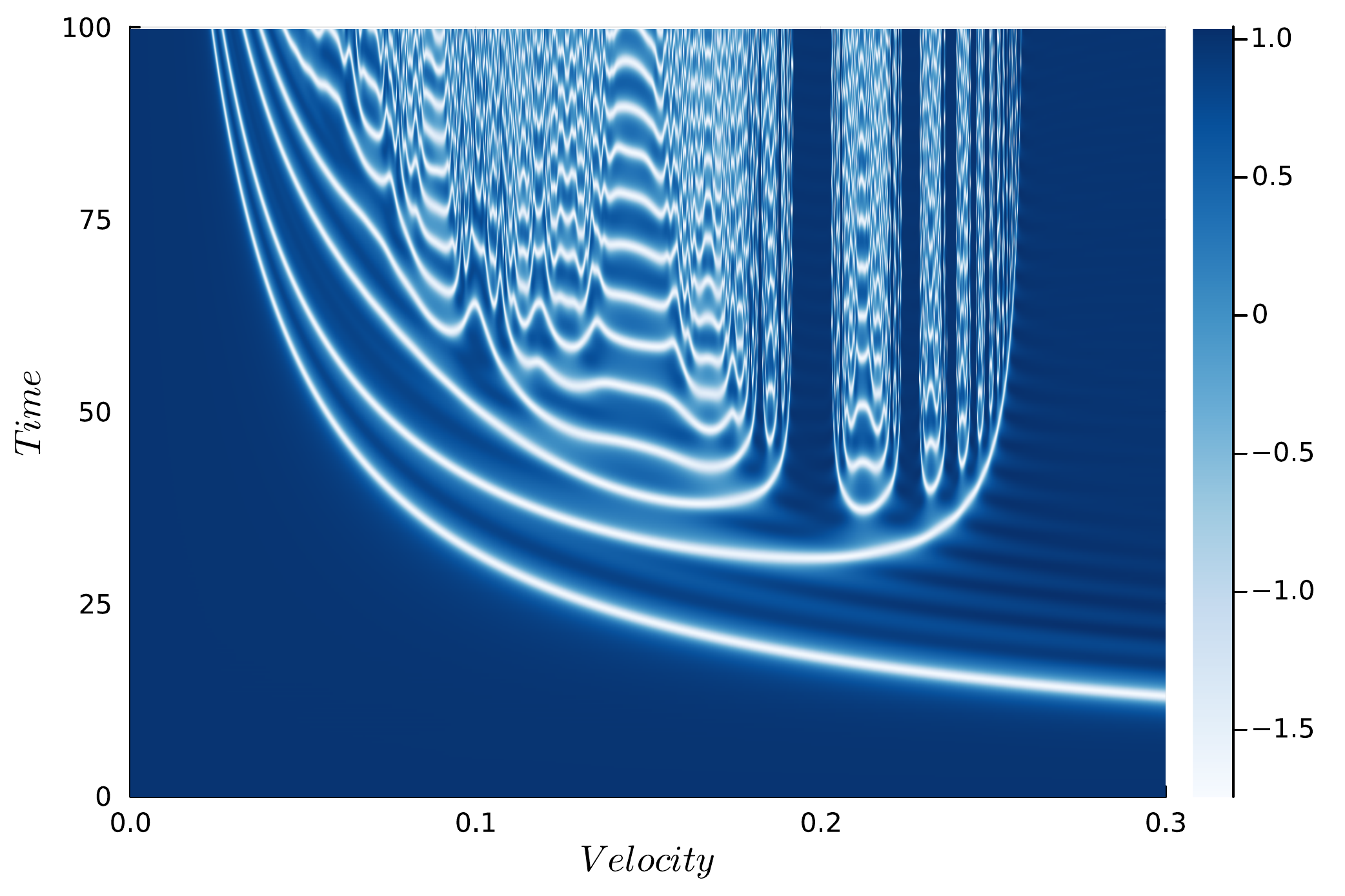}
\end{center}
{
    \caption{\small Evolution of the center field value $\phi(x=0,t)$ of $K\bar{K}$ configuration for a range of initial velocities in the $\phi^4$ model.}
    \label{fig:themap}
}
\end{figure}

In Fig.~\ref{fig:themap}, we showcase the evolution of the central field value $\phi(x=0,t)$ as a function of time and initial velocity of the $K\bar{K}$ configuration in the $\phi^4$ model. This picture demonstrates the intricate dependence of the collision's outcome on the initial velocity. More precisely, we see that the bouncing happens only in certain windows that occur below the critical velocity $v_{\rm crit} \approx 0.26$ and above $v_{\rm min} \approx 0.18$. In between the bouncing windows there are the so-called `bion chimneys' where the $K\bar{K}$ pair form a long-living, quasi-periodic state that slowly decays via emission of radiation.

Both quantitative and qualitative understanding of this phenomenology is commonly pursued through the so-called Collective Coordinate Models (CCMs).
This approach aims to reduce the infinitely-dimensional dynamics of the field theory down to a few most relevant degrees of freedom. The strategy is to select a background ansatz -- a continuous family of curves $\phi_{\rm bkg} (x; \{X_a(t)\})$ controlled by a given number of parameters $X_a$ that may vary with time. 
For a relativistic field theory with a single scalar field, i.e.
\begin{equation}
\mathcal{L} = \frac{1}{2}\partial_\mu\phi\partial^\mu\phi -V(\phi)\,,
\end{equation}
the effective Lagrangian has a generic structure
\begin{equation}
L_{\rm eff}=  \frac{1}{2}g_{ab}\dot X_a \dot X_b - U(X)\,,
\quad a,b \in \{1,\ldots\, N\}
\end{equation}
where $N$ is the number of collective coordinates and where the metric and the potential are given by the integrals
\begin{gather}
g_{ab} \equiv \int\limits_{-\infty}^{\infty}\diff x\, \frac{\partial \phi_{\rm bkg}}{\partial X_a} \frac{\partial \phi_{\rm bkg}}{\partial X_b}\,, \\
U(X) \equiv \int\limits_{-\infty}^{\infty}\diff x\, \biggl( \frac{1}{2} \phi_{\rm bkg}^{\prime\, 2} + V(\phi_{\rm bkg})\biggr)\,.
\end{gather}

The utility of CCM encoded in $L_{\rm eff}$ depends very sensitively on $\phi_{\rm bkg}$. 
Regarding the strategies for selecting viable ansatzes, we may postulate two complementary philosophies: i) the \emph{engineering} approach and ii) the \emph{agnostic} approach.

The engineering approach -- as the name suggests -- relies on incorporating prior information into the $\phi_{\rm bkg}$. If the goal is the analysis of $K\bar{K}$ scattering, for instance, the ansatz typically consists of a superposition of kink and anti-kink solutions  plus a selected number of normal modes -- a route that have been applied for, e.g. $\phi^4$ model \cite{Manton:2020onl, Manton:2021ipk} (see \cite{2019arXiv190903128K} for the somewhat intricate history of its deployment). However, CCMs that has been proposed also include Derrick modes \cite{Adam:2021gat}, quasi-normal modes \cite{Dorey:2017dsn} and/or delocalized modes \cite{Adam:2022mmm, Adam:2019xuc, Adam:2022bus}. 
In fact, the engineering approach has become a precision tool for predicting major features of the $K\bar{K}$ scattering, such as the critical velocity \cite{Adam:2023qgx}. 

Despite its successes, the engineering approach has also disadvantages. In this approach, a given CCM is like a microscope that has been carefully trained on a particular spot of the sample. Regardless how successfully a CCM models the selected feature of dynamics it has no direct applicability on other aspects. Neither it can be used for discovering new dynamical features nor to unearth connections between known ones. In short, an engineering CCM is, by construction, a single-purpose tool. 

On the other hand, the agnostic approach aims to be a general-purpose tool. Rather than carefully constraining the field(s) into a premeditated straitjacket, the agnostic CCM attempts to capture rough features of the field dynamics on a coarse-grained canvas. In this regard, it is mainly a tool for exploration. In practice, the best approach is a judicious synthesis of the two: deployment of agnostic CCM should be followed by an engineering one. Indeed, the findings of the former can be a posteriori verified and developed by the latter. Ideally, such a combo may allow exhaustive exploration of soliton dynamics in situations, where  there is a vast space of initial configurations involving multiple fields and a higher number of spatial dimensions which makes numerical solutions of field theory very time-consuming. 

To achieve this, we must first develop a toolkit for agnostic CCMs in various field theories, starting with a single scalar field in (1+1)-dimensions.  An agnostic CCM must be exhaustive, meaning that the background ansatz $\phi_{\rm bkg}$ approaches the continuum field in the limit $N\to \infty$. Furthermore, it must be algebraically tractable -- the number of terms in the effective Lagrangian should grow linearly with $N$. This is to ensure that stepping from $N$ to $N+1$ does not not generate exponential increase in complexity. 

In our previous paper \cite{Blaschke:2022fxp}, we proposed an early candidate for such an agnostic CCM that we have dubbed \emph{mechanization}. The idea is to replace a continuum field with a piece-wise linear function -- a \emph{mech-field}. We have cataloged basic features of mech-field dynamics for a few lowest values of $N$ -- which is the number of non-flat segments connected by $N+1$ joints. 

The most apparent advantage of the mechanization procedure is that it allows progressive exploration of the dynamics. As $N$ increases, more modes of behavior becomes possible. 

At $N=1$, the mech-field is a mechanical analog of the kink -- a \emph{mech-kink} (see Fig.~\ref{fig:three}). Let us point out two of its salient features: i) a static mech-kink can be boosted, despite the explicit breakdown of the Lorentz invariance that is typical for most CCM's and ii) the mech-kink has an exact periodic solution -- the so-called Derrick mode. In fact, the structure of the effective Lagrangian turns out to be virtually identical to a field-theoretical relativistic CCM for a kink \cite{Adam:2021gat}.

At $N=2$, the mech-field connecting the same vacua behaves as a quasi-periodic oscillator that can decay -- the joints fly to opposite infinities while the mech-field settles on the vacuum exponentially fast. In \cite{Blaschke:2022fxp}, we investigated how the lifetime of  this \emph{mech-oscillon} depends on its initial dimensions. More importantly, we have shown that higher-$N$ mech-oscillons can decay via multiple channels, including disintegration into excited pair of mech-kink and anti-mech-kink that -- before escaping to infinity -- may undergo several bounces.

Although our findings were encouraging, we have also identified several shortcomings of mechanization as proposed in \cite{Blaschke:2022fxp}. For example, the moduli space of a generic mech-field turned out to be geodetically incomplete, having multiple singularities corresponding to situations when joints overlap. Further, we have also encountered a technical issue that prevented us from direct investigations of mech-$K\bar{K}$ scattering. Because the segment between a mech-$K\bar{K}$ pair lies precisely in a vacuum, there is no force between them, unlike in the field theory, where a short-range attractive force exists due to overlapping tails of kinks. Thus, a direct scattering of mech-kinks seemed to be impossible, while scattering of approximate mech-kinks turned out to be riddled by numerical instabilities and presence of long-range forces.

In this paper, we present a solution to the above issues in addition to other conceptual advancements. Hence, we provide a significant step towards the construction of truly general-purpose CCM.  

Our main findings are distributed in the paper as follows. In Sec.~\ref{sec:II}, we reintroduce the mechanization procedure and provide explicit formulas for the effective Lagrangian using two different sets of coordinates. More importantly, we define a concept of \emph{BPS mechanization} that allows the construction of Bogomol'nyi-Prasad-Sommerfeld equations for static mech-kinks  saturating the same Bogomol'nyi bound  \cite{Bogomolny:1975de}  as field-theoretical kinks. 

In Sec.~\ref{sec:III}, we compare properties of mech-kinks based on non-BPS and BPS mechanization, including the discussion of normal modes. 

Sec.~\ref{sec:IV} contains an investigation of direct mech-$K\bar{K}$ scatterings for the simplest mech-fields. We first present a resolution of the decoupling problem: this is accomplished via \emph{LOse Order Mechanization}, or LOOM. In short, we show that short-range interactions of kinks in a field theory are replaced by \emph{contact} interactions between mech-kinks. By allowing the joints to pass through each other (without encountering any singularities) we continue the free dynamics of a mech-$K\bar{K}$ pair into to different `stage', where it becomes a mech-oscillon. This mech-oscillon may either decay or again form a new mech-$K\bar{K}$ pair, which can fly apart or undergo bouncing. In this way, we show that both key features, namely bouncing and (mech-)bion formation, are represented even in the simplest mech-$K\bar{K}$ scatterings. We showcase numerical results for both non-BPS and BPS kinks in $\phi^4$ model. 

Lastly, in Sec.~\ref{sec:V} we discuss the presented results and point out the future directions for the mechanization program.

\section{Mechanization}
\label{sec:II}

In this section, we gather all the technical aspects of the mechanization procedure; we define the mech-field and discuss associated moduli space providing explicit formulas for the metric via two complementary choices of coordinates. Lastly, we provide an explicit form for the effective Lagrangian for both non-BPS and BPS approaches.

\subsection{Mech-field}

\begin{figure*}[htb!]
\begin{center}
\includegraphics[width=0.9\textwidth]{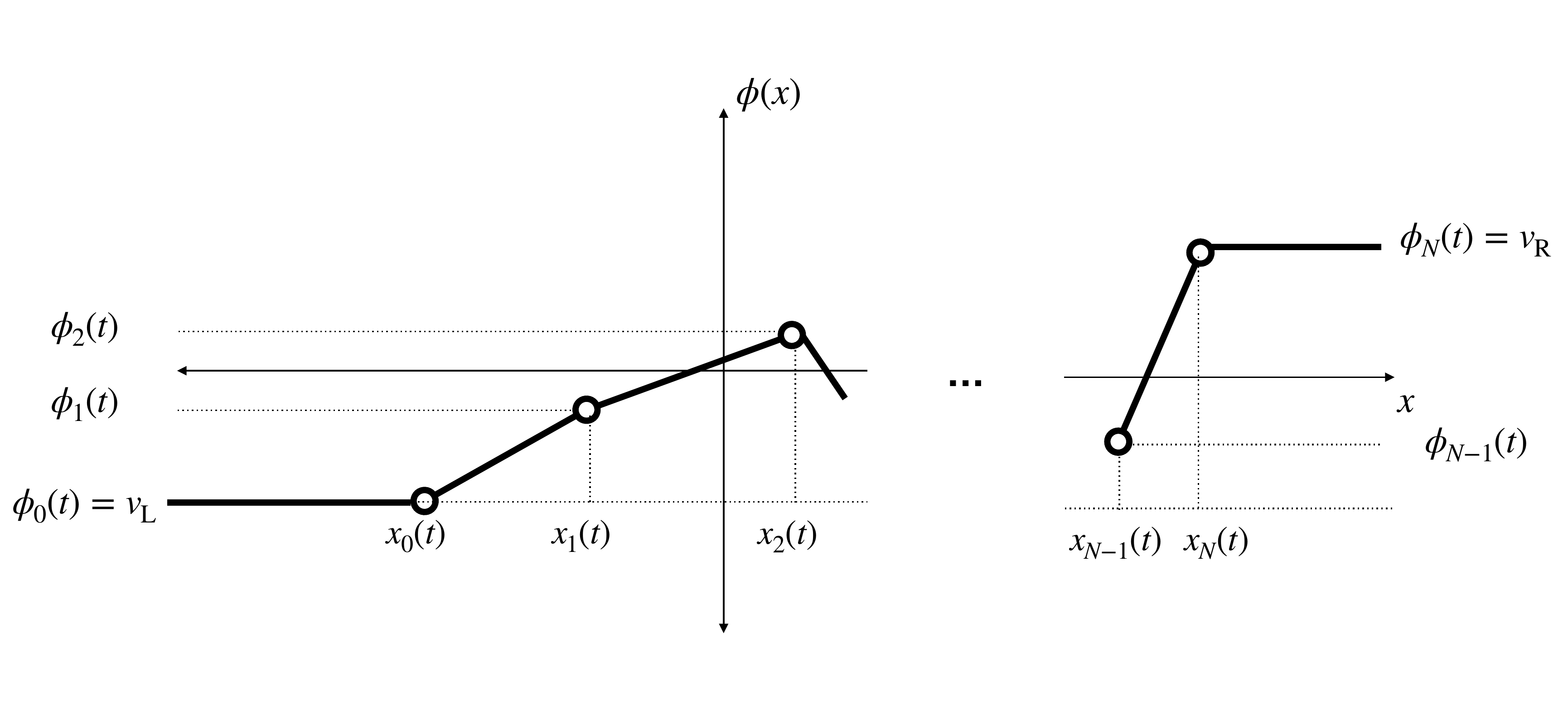}
\end{center}
{
    \caption{\small A depiction of a mech-field $\phi_M(x,t)$ as a sequence of $N$ straight stretchable segments connected via massless joints.}
    \label{fig:one}
}
\end{figure*}

The mechanization procedure replaces a continuous field $\phi(x,t)$ by a piece-wise linear function that is defined by a set of $N+1$ control points (or \emph{joints}) in the $\phi\mbox{-}x$ plane, i.e. $\{x_a, \phi_a\}$, $a = 0\, \ldots\, N$ (see Fig.~\ref{fig:one}). 
We define a \emph{mech-field} $\phi_M(x,t)$  by the formula
\begin{equation}\label{eq:decons2}
\phi_M(x,t) \equiv \sum\limits_{a=-1}^{N}\Bigl(\frac{\Delta \phi_a(t)}{\Delta x_a(t)}(x-x_a(t))+\phi_a(t)\Bigr)\chi_a\,.
\end{equation}
Here, $\Delta f_a(t) \equiv f_{a+1}(t)-f_a(t)$ and 
the $\chi_a$'s are the indicator functions for each segment:
\begin{equation}
\chi_a \equiv \theta(x-x_a)-\theta(x-x_{a+1}) = -\Delta \theta(x-x_a)\,,
\end{equation}
where $\theta(x)$ is the Heaviside's step function, i.e. $\theta(x) = 1$ if $x>0$ and $\theta(x) = 0$ otherwise.

The $x_a(t)$'s are the positions of the joints on the $x$-axis. Note that the $\phi_a$'s correspond to the values of the field at $a$-th joint, i.e. $\phi_a(t)\equiv \phi(x_a(t))$, \emph{only} if they are canonically ordered, namely  $x_0(t)< x_1(t)< \ldots < x_N(t)$. Throughout this section, we assume that this ordering holds.

We impose boundary conditions on a mech-field so that it has finite energy. Namely, we fix the two outermost segments in some vacua, i.e. $\phi_{-1} = \phi_0  =v_{\rm L}$ and $\phi_{N} = \phi_{N+1} = v_{\rm R}$ where $v_{\rm L, R}$ represent vacuum values on the left or right, respectively. We have formally added two static joints at spatial infinities, namely  $x_{-1} = -\infty$ and $x_{N+1} = +\infty$. The continuity of the mech-field can be then verified by direct differentiation
\begin{equation}
\partial_x \phi_M(x,t) = \sum\limits_{a=-1}^{N}\frac{\Delta \phi_a(t)}{\Delta x_a(t)}\chi_a-  \sum\limits_{a=-1}^{N}\Delta \bigl(\delta(x-x_a)\phi_a\bigr)\,.
\end{equation}
The second term on the r.h.s vanishes due to `fundamental theorem of discrete calculus', i.e. 
{\small \begin{multline*}
\sum\limits_{a=-1}^{N}\Delta \bigl(\delta(x-x_a)\phi_a\bigr) = \phi_{N+1}\delta(x-x_{N+1})-\phi_{-1}\delta(x-x_{-1}) \\
=v_{\rm R}\delta(x-\infty) -v_{\rm L}\delta(x+\infty) = v_{\rm R}\delta(-\infty)-v_{\rm L}\delta(\infty) = 0\,.
\end{multline*}}
A similar argument can be made to show that $\partial_t \phi_M(x,t)$ is free of delta-functions too.

There are $N-1$ `heights' of joints $\phi_1, \ldots, \phi_{N-1}$ together with $N+1$ positions $x_0, \ldots, x_N$ totalling $2N$ degrees of freedom to describe a mech-field with $N+1$ joints.

For further purposes, let us also introduce an alternative parametrization:
\begin{equation}\label{eq:dual}
\phi_M(x,t) \equiv \sum\limits_{a=-1}^{N}\Bigl(k_a(t) x+\Phi_a(t)\Bigr)\chi_a\,.
\end{equation}
Here, the $k_a$'s are the slopes of the segments, i.e.\footnote{Here, the notation is slightly different from our previous paper \cite{Blaschke:2022fxp}, where we wrote $k_{a+1}$ instead.}
\begin{equation}\label{eq:dual1}
k_{a} \equiv \frac{\phi_{a+1}-\phi_a}{x_{a+1}-x_a}\,,
\end{equation}
while the $\Phi_a$'s are given as
\begin{equation}\label{eq:dual2}
\Phi_a \equiv \frac{x_{a+1}\phi_a - x_a \phi_{a+1}}{x_{a+1}-x_a}\,.
\end{equation}

The boundary conditions reads
\begin{equation}
k_{-1} = k_{N} = 0\,, \hspace{5mm}
\Phi_{-1} = v_{\rm L}\,, \hspace{3mm}
\Phi_{N} = v_{\rm R}\,.
\end{equation}
The inverse formulas to \refer{eq:dual1}-\refer{eq:dual2} are given as
{\small \begin{equation}\label{eq:inverse}
x_{a+1} = - \frac{\Phi_{a+1}-\Phi_a}{k_{a+1}-k_a}\,, \hspace{5mm}
\phi_{a+1} = \frac{k_{a+1}\Phi_a -k_a \Phi_{a+1}}{k_{a+1}-k_a}\,.
\end{equation}}

\begin{figure}[htb!]
\begin{center}
\includegraphics[width=0.99\columnwidth]{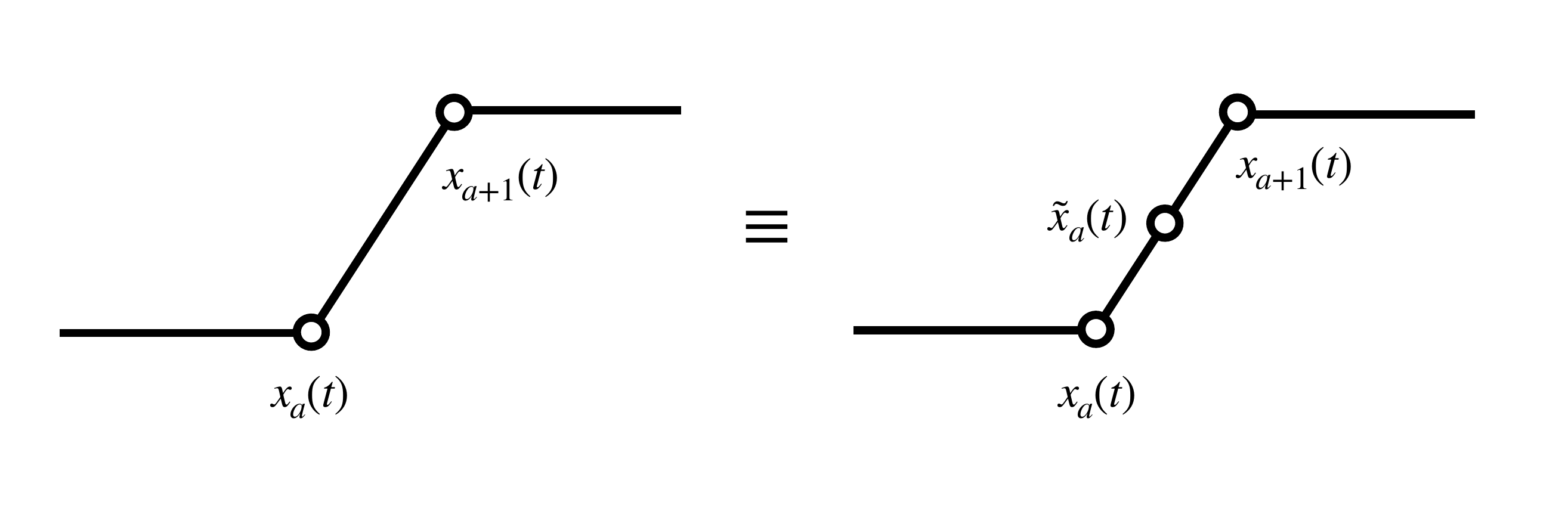}
\end{center}
{
    \caption{\small Illustration of a degeneracy of $\{x_a(t), \phi_a(t)\}$ coordinates: insertion of a new joint on any segment does not change the mech-field.}
    \label{fig:two}
}
\end{figure}

The $\{k_a, \Phi_a\}$ coordinates offer some advantages over $\{x_a, \phi_a\}$. For example, the metric, discussed in the next subsection, has the simplest form. A more subtle issue is the redundancy (or degeneracy) of $\{x_a,\phi_a\}$ coordinates. We illustrate this in Fig.~\ref{fig:two}: if we artificially add a joint on any segment, while keeping the neighboring slopes the same, the mech-field does not change, i.e. the new joint is not dynamical.
In particular, a vacuum configuration, i.e. $\phi_M = v$, can be described with a single segment, two segments, or any number of segments, with  the positions $\{x_a\}$ undetermined by the dynamics for any $N$. This can be seen directly from the formula \refer{eq:decons2} by setting $\phi_0 = \ldots = \phi_{N} = v$. 

In  $\{k_a, \Phi_a\}$ coordinates, on the other hand, the vacuum is given by $k_0 = \ldots = k_N = 0$ and $\Phi_0 = \ldots = \Phi_N = v$ and there are no undetermined degrees of freedom. Furthermore, there is truly only a single segment, because whenever two subsequent $\Phi_a$'s and $k_a$'s equal each other, the $x_{a}$ is undefined through Eq.~\refer{eq:inverse}. This is most easily seen from the rewriting of \refer{eq:dual} as
{\small \begin{equation}
\phi_M(x,t) \equiv v_{\rm L} + \sum\limits_{a=0}^{N}\theta(x-x_a(t))\Bigl(\Delta k_a(t) x+ \Delta \Phi_a(t)\Bigr)\,,
\end{equation}}
that shows that whenever $\Delta k_a = \Delta \Phi_a = 0$, the coordinate $x_a$ disappears. 

\subsection{Moduli space}

Generically, for any set of collective coordinates $\{X_a\}$ the metric is given as
\begin{equation}
g(\{X\})_{ab} \equiv \int\limits_{-\infty}^{\infty}\diff x\, \frac{\partial \phi}{\partial X_a}\frac{\partial \phi}{\partial X_b}\,.
\end{equation}

In the  $\{x_a,\phi_a\}$ coordinates, the metric consists of $(N+1)\times (N+1)$, $(N+1)\times (N-1)$ and $(N-1)\times (N-1)$ \emph{tri-diagonal} blocks, namely:
\begin{equation}
g(\{x,\phi\}) = 
\begin{pmatrix}
g^{xx} & g^{x\phi} \\
g^{\phi x} & g^{\phi\phi}
\end{pmatrix}\,,
\end{equation}
where $g^{\phi x} = (g^{x \phi})^T$ and 
{\small \begin{equation}
g^{xx} = \begin{pmatrix}
\frac{(\Delta \phi_0)^2}{3\Delta x_0} & \frac{(\Delta \phi_0)^2}{6\Delta x_0} & 0 & \ldots \\
 \frac{(\Delta \phi_0)^2}{6\Delta x_0} &  \frac{(\Delta \phi_0)^2}{3\Delta x_0}+ \frac{(\Delta \phi_1)^2}{3\Delta x_1} & \frac{(\Delta \phi_1)^2}{6\Delta x_1}& \ldots  \\
 0 & \frac{(\Delta \phi_1)^2}{6\Delta x_1} &  \frac{(\Delta \phi_1)^2}{3\Delta x_1}+ \frac{(\Delta \phi_2)^2}{3\Delta x_2} & \ldots \\
 \vdots & \vdots & \vdots & \ddots 
\end{pmatrix}
\end{equation}}

{\small \begin{equation}
g^{x\phi} = \begin{pmatrix}
(\phi_0-\phi_1)/6 & 0 & 0 & \ldots \\
(\phi_0-\phi_2)/3 & (\phi_1-\phi_2)/6 & 0 & \ldots \\
(\phi_1-\phi_2)/6 & (\phi_1-\phi_3)/3 & (\phi_2-\phi_3)/6 & \ldots \\
0 & (\phi_2-\phi_3)/6 & (\phi_2-\phi_4)/3 & \ldots \\
\vdots & \vdots & \vdots & \ddots
\end{pmatrix}
\end{equation}}

{\small \begin{equation}
g^{\phi \phi} = \begin{pmatrix}
 (x_2-x_0)/3 & (x_2-x_1)/6 & 0 & \ldots \\
 (x_2-x_1)/6 & (x_3-x_1)/3 & (x_3-x_2)/6 & \ldots \\
 0 & (x_3-x_2)/6 & (x_4-x_2)/3 & \ldots \\
\vdots & \vdots & \vdots & \ddots 
\end{pmatrix}
\end{equation}}

The determinant reads\footnote{Note that the formula for the determinant given in our previous paper \cite{Blaschke:2022fxp} was written incorrectly.}
{\small \begin{equation}\label{eq:corrdet}
\bigl|g(\{x,\phi\})\bigr| = \frac{1}{12^N}\prod\limits_{a=-1}^{N-1}\bigl(k_{a+1}-k_a\bigr)^2 \prod\limits_{b=0}^{N-1}(x_{b+1}-x_b)^2\,.
\end{equation}}
In these coordinates the metric is degenerate, i.e. $|g|=0$, not only when positions of neighboring joints coincide, namely $\Delta x_a =0$, but also when subsequent slopes are equal: $\Delta k_a =0$. The latter type of singularity reflects the aforementioned degeneracy. 

On the other hand, in $\{k_a, \Phi_a\}$ coordinates, the metric consists of four $N\times N$ \emph{diagonal} blocks:
\begin{equation}
g(\{k,\Phi\}) = \begin{pmatrix}
g^{kk} & g^{k \Phi} \\
g^{\Phi k} & g^{\Phi\Phi} 
\end{pmatrix}\,,
\end{equation}
where $g^{\Phi k} = g^{k \Phi}$ and
\begin{equation}
g^{kk} = \frac{1}{3}
\begin{pmatrix}
x_1^3-x_0^3 & 0 & 0 & \ldots \\
0 & x_2^3- x_1^3 & 0 & \ldots \\
0 & 0 & x_3^3-x_2^3 & \ldots \\
\vdots & \vdots & \vdots & \ddots 
\end{pmatrix}\,,
\end{equation}
\begin{equation}
g^{k\Phi} = \frac{1}{2}
\begin{pmatrix}
x_1^2-x_0^2 & 0 & 0 & \ldots \\
0 & x_2^2- x_1^2 & 0 & \ldots \\
0 & 0 & x_3^2-x_2^2 & \ldots \\
\vdots & \vdots & \vdots & \ddots 
\end{pmatrix}\,,
\end{equation}
\begin{equation}
g^{\Phi\Phi} = 
\begin{pmatrix}
x_1-x_0 & 0 & 0 & \ldots \\
0 & x_2- x_1 & 0 & \ldots \\
0 & 0 & x_3-x_2 & \ldots \\
\vdots & \vdots & \vdots & \ddots 
\end{pmatrix}\,.
\end{equation}
Furthermore, the determinant reads
\begin{equation}\label{eq:det2}
\bigl|g(\{k,\Phi\})\bigr| = \frac{1}{12^N}\prod\limits_{a=0}^{N-1}(x_{a+1}-x_a)^4\,,
\end{equation}
and contains only a singularity of the type $\Delta x_a =0$.

\subsection{The effective Lagrangian and the BPS mechanization}

To obtain an effective Lagrangian we inserts a mech-field given either by \refer{eq:decons2} or \refer{eq:dual} into a Lagrangian density $\mathcal{L}$ taken as a generic scalar-field theory in 1+1 dimensions, i.e.
\begin{equation}\label{eq:contlag}
\mathcal{L} = \frac{1}{2}{\dot\phi}^2 -\frac{1}{2}\phi^{\prime\, 2}-V(\phi)\,,
\end{equation}
and integrate it over $x$-axis
\begin{equation}
L_{\rm eff} = \int\limits_{-\infty}^{\infty} \diff x\, \mathcal{L}\bigl(\phi_M\bigr)\,.
\end{equation}
The result derived by assuming canonical ordering of joints, $x_0 < x_1 < \ldots < x_N$, reads (see the details in \cite{Blaschke:2022fxp})
{\small \begin{gather}
L\bigl[\bigl\{x, \phi\bigr\}\bigr] =  \sum\limits_{a=0}^{N-1}\Delta x_a\biggl[ \frac{1}{6}\Bigl(\Delta\dot \phi_a-\frac{\Delta \dot x_a}{\Delta x_a}\Delta\phi_a \Bigr)^2  -\frac{\Delta \phi_a^2}{2\Delta x_a^2}
\nonumber \\
+\frac{1}{2} \Bigl(\dot \phi_{a+1}-\frac{\Delta \phi_a}{\Delta x_a}\dot x_{a+1}\Bigr)\Bigl(\dot \phi_{a}-\frac{\Delta \phi_a}{\Delta x_a}\dot x_{a}\Bigr)
 - \frac{\Delta{\mathcal V}\bigl(\phi_{a}\bigr)}{\Delta \phi_{a}} \biggr]\,.
 \label{eq:efflag}
\end{gather} }
In the $\{k,\Phi\}$ coordinates, the same can be expressed slightly more compactly as follows
{\small \begin{gather}
L\bigl[\bigl\{k, \Phi\bigr\}\bigr] =  \sum\limits_{a=0}^{N-1}\biggl[ \frac{x_{a+1}^3-x_a^3}{6}\dot k_a^2 +\frac{x_{a+1}^2-x_a^2}{2}\dot k_a \dot \Phi_a + \frac{\Delta x_a}{2}\dot\Phi_a^2 
\nonumber \\
 -\frac{1}{2}k_a^2\Delta x_a
 - \Delta x_a \frac{\Delta{\mathcal V}\bigl(\phi_{a}\bigr)}{\Delta \phi_{a}} \biggr]\,,
 \label{eq:efflag2}
\end{gather} }
where $x_a$'s and $\phi_a$'s are understood as functions of $k_a$'s and $\Phi_a$'s through relations \refer{eq:inverse}.
In both formulas \refer{eq:efflag}-\refer{eq:efflag2}, $\mathcal{V}(\phi)$ is the primitive function o the potential $V(\phi)$, i.e. $\mathcal{V}^\prime(\phi) = V(\phi)$.

Let us stress that \refer{eq:efflag} and \refer{eq:efflag2} are valid only if $x_0 < x_1 < \ldots < x_N$. We will return to this point in Sec.~\ref{sec:IV}, where we present the effective Lagrangian (LOOM) that incorporates all posible orderings. 

Let us now point out that mechanization of the potential term, i.e.
\begin{equation}\label{eq:nonbpspot}
\int\limits_{-\infty}^{\infty}\diff x\, V(\phi_M) = \sum\limits_{a=0}^{N-1} \Delta x_a\frac{\Delta \mathcal{V}(\phi_a)}{\Delta \phi_a} \,,
\end{equation}
obtained by a direct integration is not unique and may not be the most optimal for studying topological solutions. In the following, let us label the outcome \refer{eq:nonbpspot} a \emph{non-BPS mechanization} for the reasons that become obvious. 

Now, let us consider a field theory in the form
\begin{equation}
\mathcal{L}_J = \frac{1}{2}\partial_\mu \phi\partial^\mu\phi +\frac{1}{2}J^2 +J W(\phi)\,,
\end{equation}
where $W(\phi)$ is the superpotential, i.e. $V(\phi) \equiv \tfrac{1}{2}W^2(\phi)$, and where $J$ is an auxiliary field. Of course, $\mathcal{L}_J$ is physically equivalent to \refer{eq:contlag} as can be seen by eliminating $J$ through its equation of motion: $J = -W(\phi)$ and plugging it back.

However, if we \emph{first} mechanize the auxiliary field as
\begin{equation}
J_M = \sum\limits_{a=0}^{N-1}\bigl(\theta(x-x_a(t))-\theta(x-x_{a+1}(t))\bigr) J_a(t)\,, 
\end{equation}
where \emph{the positions of joints $x_a(t)$ are the same as those appearing in} $\phi_M$, 
inserting both $J_M$ and $\phi_M$ into $\mathcal{L}_J$ and integrating over $x$ yields
{\small \begin{align}
L_M^{J}  \supset & \int\limits_{-\infty}^{\infty}\diff x\, \Bigl(\frac{1}{2}J_M^2 +J_M W\bigl(\phi_M\bigr)\Bigr) \nonumber \\
= & 
\sum\limits_{a=0}^{N-1}\biggl(\frac{\Delta x_a}{2}J_a^2 +\Delta x_a J_a \frac{\mathcal{W}(\phi_{a+1})-\mathcal{W}(\phi_a)}{\phi_{a+1}-\phi_a}\biggr)\,,
\end{align}}
where $\mathcal{W}$ is a primitive function of the superpotential, i.e. $\mathcal{W}^{\prime} = W$.

Eliminating all $J_a$'s via their equations of motion, we arrive at what we dub \emph{BPS mechanization}, namely:
\begin{gather}
L_M^{J} \xrightarrow[J_a = -\Delta \mathcal{W}_a/\Delta\phi_a]{}  L_{M}^{\rm BPS}  \\
L_{M}^{\rm BPS} \equiv \sum\limits_{a=0}^{N-1}\biggl[ \frac{x_{a+1}^3-x_a^3}{6}\dot k_a^2 +\frac{x_{a+1}^2-x_a^2}{2}\dot k_a \dot \Phi_a + \frac{\Delta x_a}{2}\dot\Phi_a^2 
\nonumber \\
 -\frac{1}{2}k_a^2\Delta x_a
 - \frac{\Delta x_a}{2}\biggl(\frac{\Delta{\mathcal{W}}\bigl(\phi_{a}\bigr)}{\Delta \phi_{a}}\biggr)^2 \biggr]\,.
 \end{gather}

In other words, we have found that the following loop does not close (the horizontal arrows $\xrightarrow[M]{}$ denote mechanization, while the vertical arrows denote elimination of auxiliary variables):
\begin{equation}
\begin{CD}
\mathcal{L}_J @>>M> L_M^J \\
@VV{J = -W}V @VV{J_a = - \tfrac{\Delta \mathcal{W}_a}{\Delta \phi_a}}V\\
\mathcal{L} @>>M> L_M \not = L_M^{\mathrm{BPS}}
\end{CD}
\end{equation}

Although the difference between $L_M$ and $L_M^{\rm BPS}$ is isolated only to the potential term and, at first glance, does not seem significant, we will show that it has a profound impact on the nature of static solutions and their dynamics.

\section{Static solutions} 
\label{sec:III}

In this section, we study the properties of static solutions of both $L_M$ and $L_M^{\rm BPS}$ and highlight their differences. 

\subsection{Non-BPS mech-kinks}

To find static solutions of a generic $N$ mech-field we minimize the static energy:
\begin{equation}\label{eq:statice}
E_M = \sum\limits_{a=0}^{N-1}\biggl(\frac{(\Delta \phi_a)^2}{2\Delta x_a}+ \Delta x_a \frac{\Delta \mathcal{V}(\phi_a)}{\Delta \phi_a}\biggr)\,.
\end{equation}
The coordinates of the joints (up to overall position) can be found as\footnote{Here, we assume the sequence  $\{\phi_0, \phi_1, \ldots \}$ to be monotonically increasing, i.e. the solution is a mech-kink interpolating vacua $v_{\rm L}< v_{\rm R}$.  The anti-mech-kinks would be found analogously after the appropriate insertion of absolute values inside the square roots so that $\Delta x_a>0$ for all segments.}
\begin{equation}\label{eq:pos}
\Delta x_a = \frac{(\Delta\phi_a)^{3/2}}{\sqrt{2\bigl({\mathcal V}(\phi_{a+1})-{\mathcal V}(\phi_{a})\bigr)}}\,.
\end{equation}

On the other hand, the field values $\phi_a$'s follows from minimization of \refer{eq:statice} after inserting \refer{eq:pos}, i.e.
\begin{equation}
E_M \xrightarrow[\refer{eq:pos}]{} \sum\limits_{a=0}^{N-1}\sqrt{2\Delta\phi_a \bigl({\mathcal V}(\phi_{a+1})-{\mathcal V}(\phi_{a})\bigr)}\,.
\end{equation}
This leads to a system of non-linear algebraic equations:
\begin{equation}\label{eq:fieldvals}
V(\phi_a)^2 = \frac{{\mathcal V}(\phi_{a+1})-{\mathcal V}(\phi_{a})}{\phi_{a+1}-\phi_a}\frac{{\mathcal V}(\phi_{a})-{\mathcal V}(\phi_{a-1})}{\phi_{a}-\phi_{a-1}}\,.
\end{equation}

\begin{figure}
\begin{center}
\includegraphics[width=0.95\columnwidth]{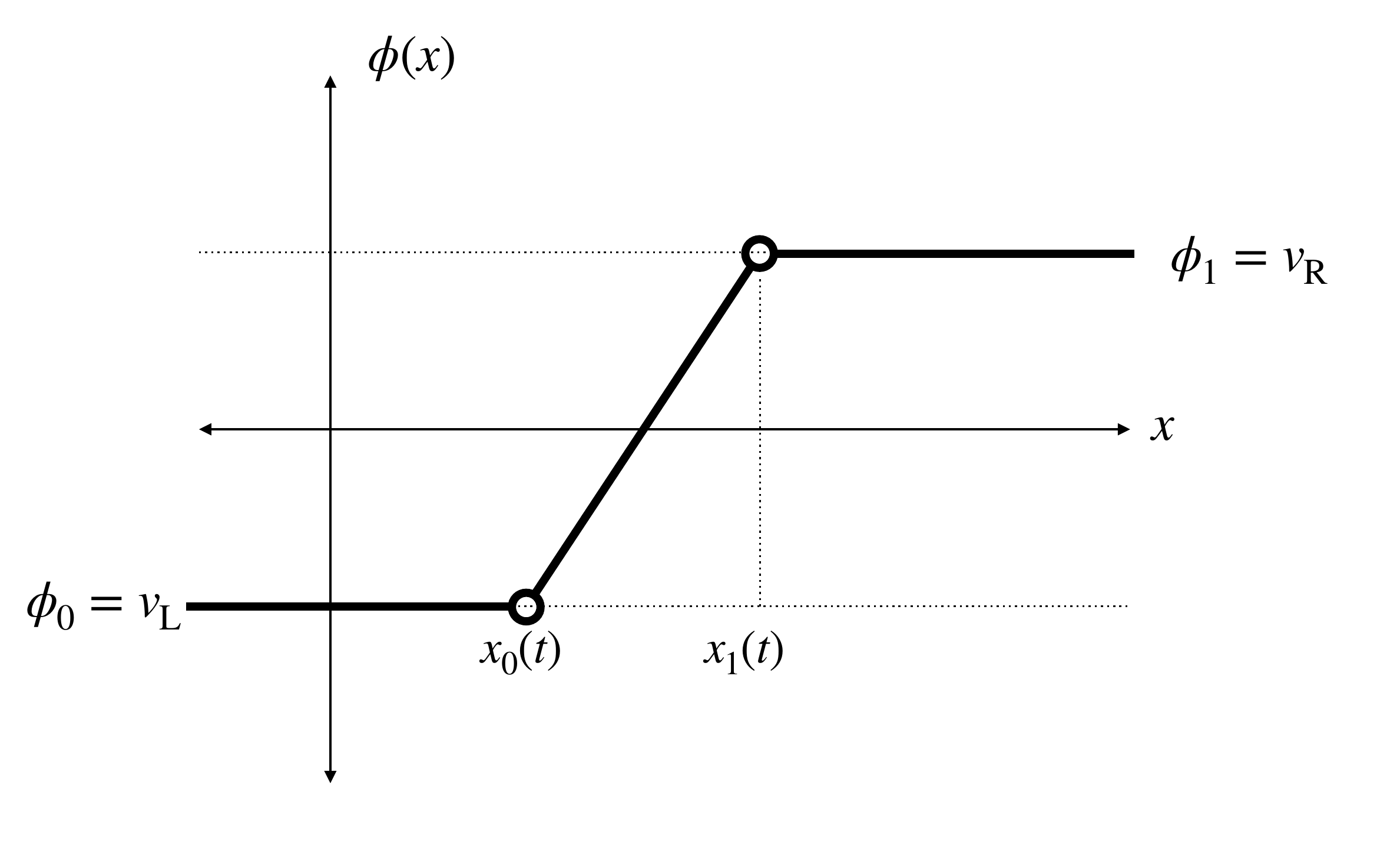}
\caption{\small Simplest mechanical model of a kink = `mech-kink'.}
\label{fig:three}
\end{center}
\end{figure}

\begin{figure}
\begin{center}
\includegraphics[width=0.95\columnwidth]{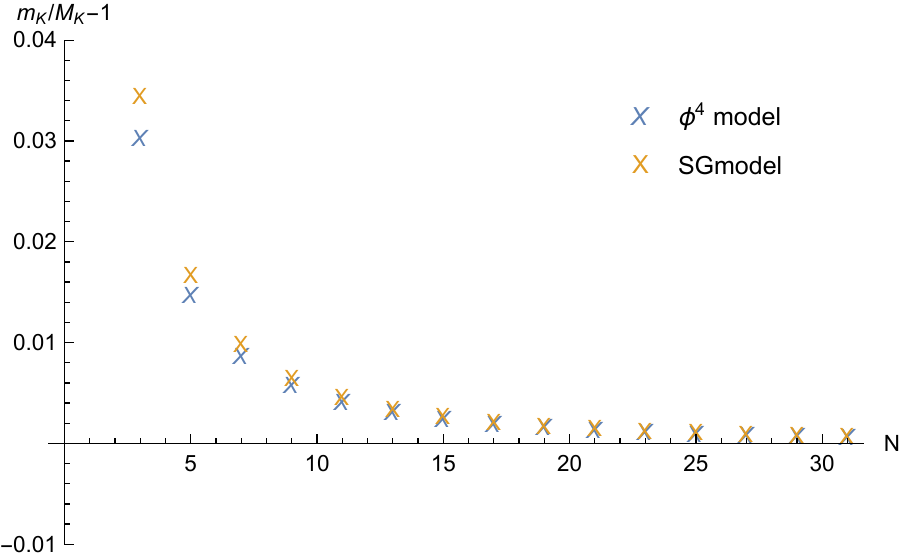}
\caption{\small The relative difference of mech-kink static energy $m_K$ and field-theoretical value $M_K$ as a function of the number of joints $N$ for $\phi^4$ and SG model.}
\label{fig:four}
\end{center}
\end{figure}

The simplest solution is $N=1$ mech-kink (see Fig.~\ref{fig:three}). Its static width $R_K$ and static energy $m_K$ are given by
\begin{equation}
R_K = \frac{v_{\rm R}-v_{\rm L}}{\sqrt{2\kappa}}\,, \hspace{5mm} 
m_K =  (v_{\rm R}-v_{\rm L})\sqrt{2\kappa}\,,
\end{equation}
where 
\begin{equation}\label{eq:kappa}
\kappa = \frac{1}{v_{\rm R}-v_{\rm L}}\int\limits_{v_{\rm L}}^{v_{\rm R}}\diff t\, V(t)\,.
\end{equation}
The corresponding values for $\phi^4$ model are $R_K = \sqrt{15/2}$ and $m_K = \sqrt{32/15} \approx 1.46$. The latter value is not that far from the field-theoretical value $M_K=4/3$. 
However, as we show in Fig.~\ref{fig:four}, the mass of higher-$N$ mech-kinks approaches $M_K$  only relatively slowly.
 
The $N=1$ mech-kink has only one massive normal mode: the Derrick mode. This mode is associated with infinitesimal scaling and has been identified as a crucial element in constructing relativistic CCM's for $K\bar{K}$ collisions in various field theories \cite{Adam:2021gat}. Its (angular) frequency is universally given as $\omega_D^2 = Q/M$, where $Q$ is the second moment of static energy density. The corresponding formula for $N=1$ mech-field reads
$q_M = (v_{\rm R}-v_{\rm L})^2/(12 R_K)$.
In the case of $\phi^4$ theory, we have $q_M = \sqrt{5/6}\approx 0.91$ which is quite far from field-theoretical value  $Q \approx 0.43$.

In general, a mech-kink with $N+1$ joints has $2N$ normal modes. The lowest one is a zero mode corresponding to the overall translation of joints. The remaining $2N-1$ modes are massive modes. As $N\to \infty$, we should somehow recover the corresponding spectrum of field-theoretical kinks. This typically consists of a certain number of localized modes and a continuous spectrum of radiation modes, depending on the model at hand. 

\begin{figure}
\begin{center}
\includegraphics[width=0.95\columnwidth]{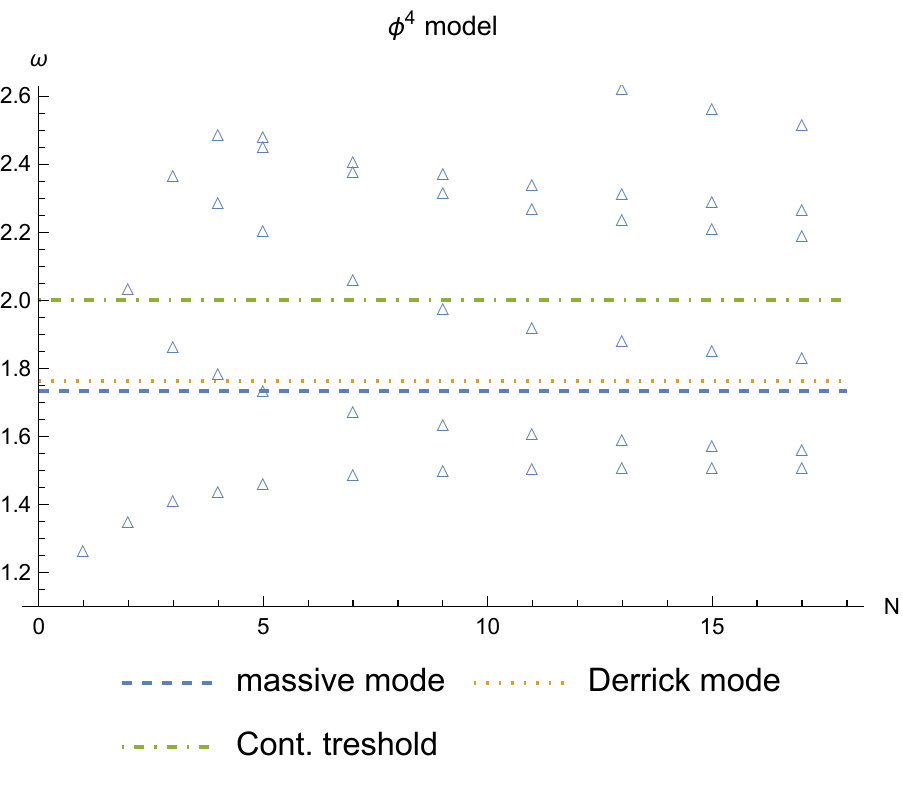}
\caption{\small Distribution of frequencies $\omega$ of normal modes as a function of $N$ for $\phi^4$ model. Only a few of the lowest frequencies are displayed and zero modes are omitted.}
\label{fig:five}
\end{center}
\end{figure}

\begin{figure}
\begin{center}
\includegraphics[width=0.95\columnwidth]{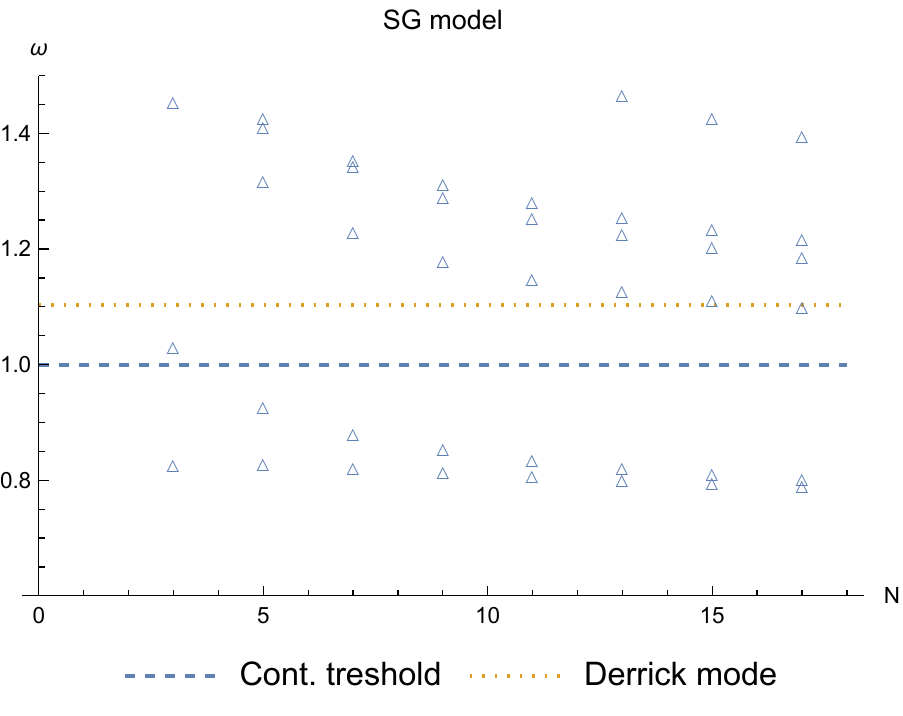}
\caption{\small Distribution of frequencies $\omega$ of normal modes as a function of $N$ for SG model. Only a few of the lowest frequencies are displayed and zero modes are omitted.}
\label{fig:six}
\end{center}
\end{figure}

In Figs.~\ref{fig:five}-\ref{fig:six}, we show that this correspondence (if it exists at all) is not very visible at the displayed range $N\leq 17$. For instance, Fig.~\ref{fig:six} hints at some convergence of the $3^{\rm rd}$ normal modes towards the (blue, dashed) line of the $\phi^4$ kink's only massive mode, but this could be entirely coincidental and further investigation into higher $N$'s is needed to draw any conclusions.

These results illustrate that the correspondence between relatively high $N$ non-BPS mech-kinks and field-theoretical kinks is not as simple, as one might hope, especially regarding the structure of normal modes.
Let us now see how the situation differs for static solutions of $L_M^{\rm BPS}$.

\subsection{BPS mech-kinks}

In the BPS scheme, the static energy reads
\begin{equation}\label{eq:statice2}
E_M^{\rm BPS} = \sum\limits_{a=0}^{N-1}\biggl(\frac{(\Delta \phi_a)^2}{2\Delta x_a}+ \frac{\Delta x_a}{2} \biggl(\frac{\Delta \mathcal{W}(\phi_a)}{\Delta \phi_a}\biggr)^2\biggr)\,.
\end{equation}
Proceeding as in the previous subsection, we first eliminate the coordinates of the joints via their equations of motion:
\begin{equation}\label{eq:pos2}
\Delta x_a = \frac{(\Delta \phi_a)^2}{\Delta \mathcal{W}(\phi_a)}\,.
\end{equation}
In contrast with the non-BPS case, if we insert this relation back into the $E_M^{\rm BPS}$ we obtain a pure number
\begin{align}
E_M^{\rm BPS} \xrightarrow[\refer{eq:pos2}]{} & \sum\limits_{a=0}^{N-1}\Delta \mathcal{W}(\phi_a) = \mathcal{W}(\phi_N)-\mathcal{W}(\phi_0) \nonumber \\
 & = \mathcal{W}(v_{\rm R})-\mathcal{W}(v_{\rm L}) \equiv M_K
\end{align}
which is given by the difference of superpotentials evaluated for vacua at $\pm \infty$ --  the field-theoretical BPS mass of the kink $M_K$!

We can establish this result in a standard way by completing the energy into a (sum of) square(s) \`a la Bogomol'nyi \cite{Bogomolny:1975de}:
\begin{align}
E_M^{\rm BPS} = & \sum\limits_{a=0}^{N-1}\biggl(\frac{(\Delta \phi_a)^2}{2\Delta x_a}+ \frac{\Delta x_a}{2} \Bigl(\frac{\Delta \mathcal{W}(\phi_a)}{\Delta \phi_a}\Bigr)^2\biggr)\nonumber \\
= & \sum\limits_{a=0}^{N-1}\frac{\Delta x_a}{2}\biggl(\frac{\Delta \phi_a}{\Delta x_a} - \frac{\Delta \mathcal{W}(\phi_a)}{\Delta\phi_a}\biggr)^2 \nonumber \\ & + \sum\limits_{a=0}^{N-1} \Delta \mathcal{W}(\phi_a) 
\geq  \sum\limits_{a=0}^{N-1} \Delta \mathcal{W}(\phi_a) = M_K\,.
\end{align}
The minimization of energy is achieved by vanishing the squares giving us the conditions \refer{eq:pos2}.

Interestingly, BPS mech-kinks have unconstrained `heights' of the joints because there is no equivalent of Eq.~\refer{eq:fieldvals} that would uniquely determine $\phi_a$'s. Indeed, $\phi_a$'s are arbitrary as long as $\Delta x_a>0$. This boils down to conditions 
\begin{equation}
\mathcal{W}(\phi_{a+1})> \mathcal{W}(\phi_a)\,, \quad \forall a\,.
\end{equation}

For $N=1$, the only difference from the non-BPS case is in the parameter $\kappa_{\rm BPS}$  (compare with Eq.~\refer{eq:kappa}):
\begin{equation}\label{eq:bpskappa}
\kappa_{\rm BPS} = \frac{1}{2(v_{\rm R}-v_{\rm L})^2}\biggl(\,\int\limits_{v_{\rm L}}^{v_{\rm R}}\diff t\, W(t)\biggr)^2\,.
\end{equation}
Its value for $\phi^4$ model is $\kappa_{\rm BPS} = 2/9$ giving us $R_K = 3$, $m_K = M_K =4/3$ and $q_M = 1$.

The structure of normal modes is also very different compared with non-BPS mech-kinks. A BPS mech-kink has $N$ zero modes (!) corresponding to an overall shift in the position of joints and freedom to make infinitesimal shifts of each $\phi_a$'s. Consequently, it has only $N$ massive normal modes in contrast with $2N-1$ massive modes for non-BPS mech-kinks.

\begin{figure}
\begin{center}
\includegraphics[width=0.95\columnwidth]{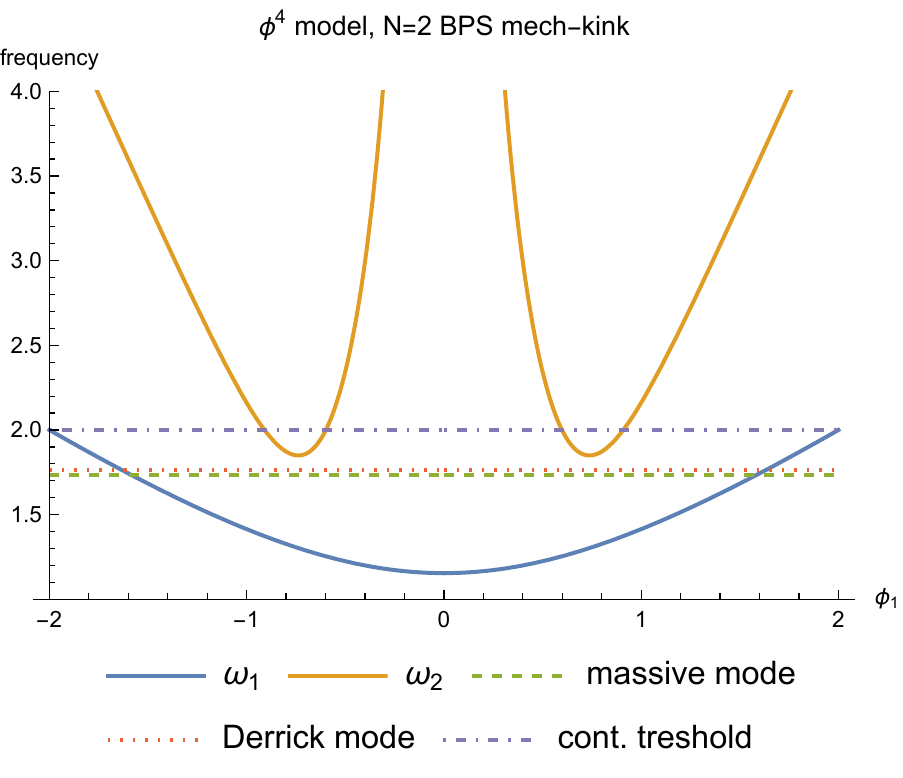}
\caption{\small Dependence of frequency of two massive modes for $N=2$ BPS mech-kink on the height of the middle joint $\phi_1$. The horizontal lines indicate relevant frequencies for field-theoretical kink in $\phi^4$ theory.}
\label{fig:smile}
\end{center}
\end{figure}

The frequencies of these massive modes, however, do depend on values of $\phi_a$'s. As an illustration, in Fig.~\ref{fig:smile} we display the frequencies of the two massive modes as functions of $\phi_1$ for the $\phi^4$ model. In fact, it is easy to work out explicit formulas:
\begin{equation}
\omega_1 = \frac{\sqrt{4+2\phi_1^2}}{\sqrt{3}}\,, \hspace{5mm}
\omega_2 = \frac{1}{3}\sqrt{54 \phi_1^2+\frac{16}{\phi_1^2}-28}\,.
\end{equation}

A salient feature of Fig.~\ref{fig:smile} is the expected mirror symmetry under $\phi_1 \to -\phi_1$ and the fact that, at the center $\phi_1=0$, the second massive mode diverges, i.e. $\omega_2 \to \infty$. This is a footprint of the coordinate degeneracy: at that point the $N=2$ BPS mech-kink is indistinguishable from $N=1$ mech-kink. Indeed, the value $\omega_1(\phi_1 =0) = 2/\sqrt{3}$ is the same as the $N=1$ Derrick mode. What is more surprising is the fact that the lengths of segments
\begin{equation}
\Delta x_0 = \frac{3}{2-\phi_1}\,, \hspace{5mm}
\Delta x_1 = \frac{3}{2+\phi_1}\,,
\end{equation}
are both positive in the range $\phi_1 \in (-2,2)$. Thus, the middle joint can also be placed outside of $[-1,1]$ contrary to expectations. 

Lastly, let us address the question of recovering the spectrum of normal modes of $\phi^4$ kink in the limit $N\to \infty$. Compared with non-BPS mech-kinks, the situation is complicated by the fact that $N$ massive normal modes depend on $N-1$ free parameters, the $\phi_a$'s.

\begin{figure}
\begin{center}
\includegraphics[width=0.95\columnwidth]{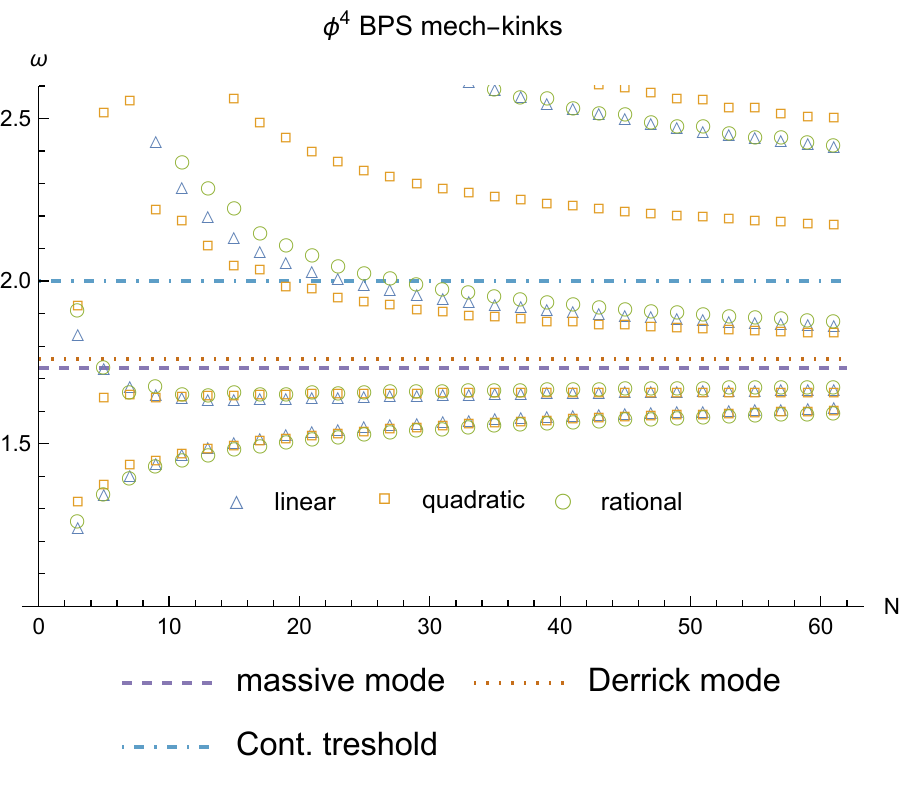}
\caption{\small Frequencies of normal modes for $\phi^4$ BPS mech-kins as a function of the number of joints $N+1$. We depict a gradual approach of the values for three different value assignments of field-values $\phi_a$'s as described in the text.}
\label{fig:convspec}
\end{center}
\end{figure}

To make progress, we studied several somewhat random types of value assignments for the heights of joints, that may be called \emph{linear}, \emph{quadratic} and \emph{rational}, given by formulae:
\begin{align}
\phi_a = & -1+ \frac{2a}{N}\,, & & \mbox{linear} \\
\phi_a = & -1+\frac{1}{2}\Bigl(\frac{2a}{N}\Bigr)^2\,, & & \mbox{quadratic} \\
\phi_a = & \phantom{-}1- \frac{2(N-a)}{a+N}\,. & & \mbox{rational}
\end{align}
In Fig.~\ref{fig:convspec}, we display how the lowest-lying frequencies of normal modes change with increasing $N$ for all three types of assignments. We see that -- especially for the two lowest massive modes -- the frequencies tend to converge to the same values for all three assignments. This hints that the same limiting spectrum should be reached for any choice of $\phi_a$'s. However, as was the case for non-BPS mech-kinks, the convergence to  $\sqrt{3}$ -- the frequency of the $\phi^4$ kink's shape mode -- is very slow if it exists at all. All that we can claim is that the convergence towards the shape mode is very slow in the displayed range $N \leq 61$.

Another crucial property of BPS mech-kinks is the existence of $N$ zero modes due to the degeneracy of the solution. For example, $N=2$ mech-kink has not only translational zero mode but also a zero mode corresponding to infinitesimal change of $\phi_1$, the height of the middle joint. 
This extra zero mode has an interesting and counter-intuitive impact on the dynamic of $N=2$ mech-kinks. We can easily imagine (and we observe it in numerical simulations) that during dynamical evolution, the coordinate $\phi_1$ drifts from its initial value, as its change does not cost any energy. We observe that once the middle joint becomes very close to one of the outer joints, the latter quickly accelerate to infinity. In this way, the $N=2$ mech-kink sheds one of its outer joints and becomes effectively $N=1$ mech-kink. This phenomenon is called `joint ejection' and we reported it in our previous paper \cite{Blaschke:2022fxp}.

Although the joint ejections were observed for sufficiently perturbed $N\geq 2$ non-BPS mech-kinks, in the BPS case they are practically inevitable due to the presence of extra zero modes.
This leaves only the $N=1$ mech-kink as a truly stable solution in BPS mechanization.

\section{Scattering of mech-kinks}
\label{sec:IV}

In this section, we investigate the simplest forms of scatterings between mech-kinks to showcase the second conceptual advancement of this paper: what we call a LOse Order Mechanization, or a LOOM. As we shall see, the analysis of mech-$K\bar{K}$ scattering requires including non-canonical orderings of the joints and construction of an effective Lagrangian that incorporates these different orderings. We also present numerical results for dynamics of symmetric $N=3$ mech-field and we find qualitatively similar behavior to field theory, namely that mech-$K\bar{K}$ pair undergoes bounces or form (mech-)bions. We analyze the scattering of both non-BPS and BPS mech-kinks and comment on the differences.

\subsection{Decoupling}

The compact nature of mech-fields, i.e. that they have finite extents outside of which there are only exact vacua, allow us to superpose objects, e.g. mech-kinks or mech-oscillons, without introducing any interaction between them. 
This is most easily visible at the effective Lagrangian level. Taking a generic mech-field and fixing the $a$-th segment to a vacuum, say $\phi_{a} = \phi_{a+1} = v$, while keeping $x_a$ and $x_{a+1}$ dynamical, it is easy to see that the effective Lagrangian consists of two decoupled pieces:
\begin{equation}
L_M[\phi_M] \xrightarrow[\phi_a, \phi_{a+1}\to v]{ } L_M[\phi_M^{(1)}]+ L_M[\phi_M^{(2)}]\,.
\end{equation}
In other words, there remain no interacting terms that could inform the constituent mech-fields $\phi_M^{(1,2)}$  about each other's existence.\footnote{During dynamical evolution a mech-field can pass through decoupled configuration. In such a case, however, non-zero derivatives preserves the interaction between the two parts, so the decoupling does not occur.}
Thus, the two parts evolve according to their respective dynamics as if the other piece is not there at all. This is, of course, true only \emph{until they start to overlap}, where the very description of the dynamics via effective Lagrangians \refer{eq:efflag} or \refer{eq:efflag2} is invalid. 

\begin{figure}
\begin{center}
\includegraphics[width=0.95\columnwidth]{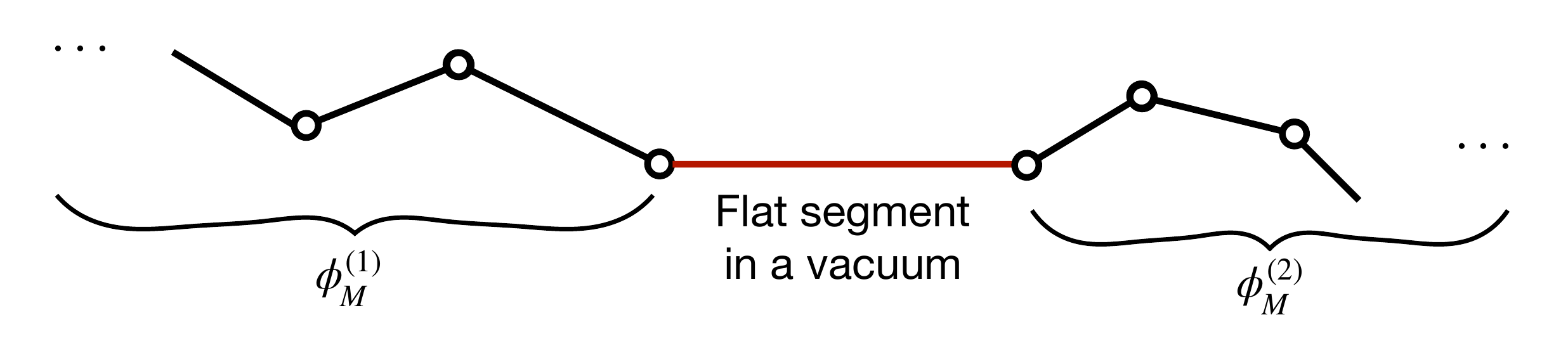}
\caption{\small Illustration of the decoupling property: If the mech-field consists of two parts connected by a flat segment in a vacuum, the coordinates in the left and right parts of the mech-field do not interact.}
\label{fig:eight}
\end{center}
\end{figure}

As a consequence, a mech-kink and an anti-mech-kink separated by a vacuum segment of arbitrary length do not impart any force on one another. This should be compared with what is going on in the field theory, where a well-separated $K\bar{K}$ configuration experiences an exponentially damped attractive force, precisely because field-theoretical solitons are not compact.\footnote{In special field theories, however, compact solutions can be present \cite{Arodz:2002yt, Arodz:2007jh, Hahne:2019ela}. Their behavior is quite similar  to mech-kinks and mech-oscillons presented here (or, more correctly, vice versa).} 
This presents somewhat of a road-block to a naive investigations of mech-$K\bar{K}$ scattering. Indeed, the dynamics derived from the effective Lagrangian \refer{eq:efflag} is trivial before the collision (no force) and undefined for the moment of contact as $L[\{x_a,\phi_a\}]$ applies only for canonically ordered mech-fields.

\begin{figure}[htb!]
\begin{center}
\includegraphics[width=0.9\columnwidth]{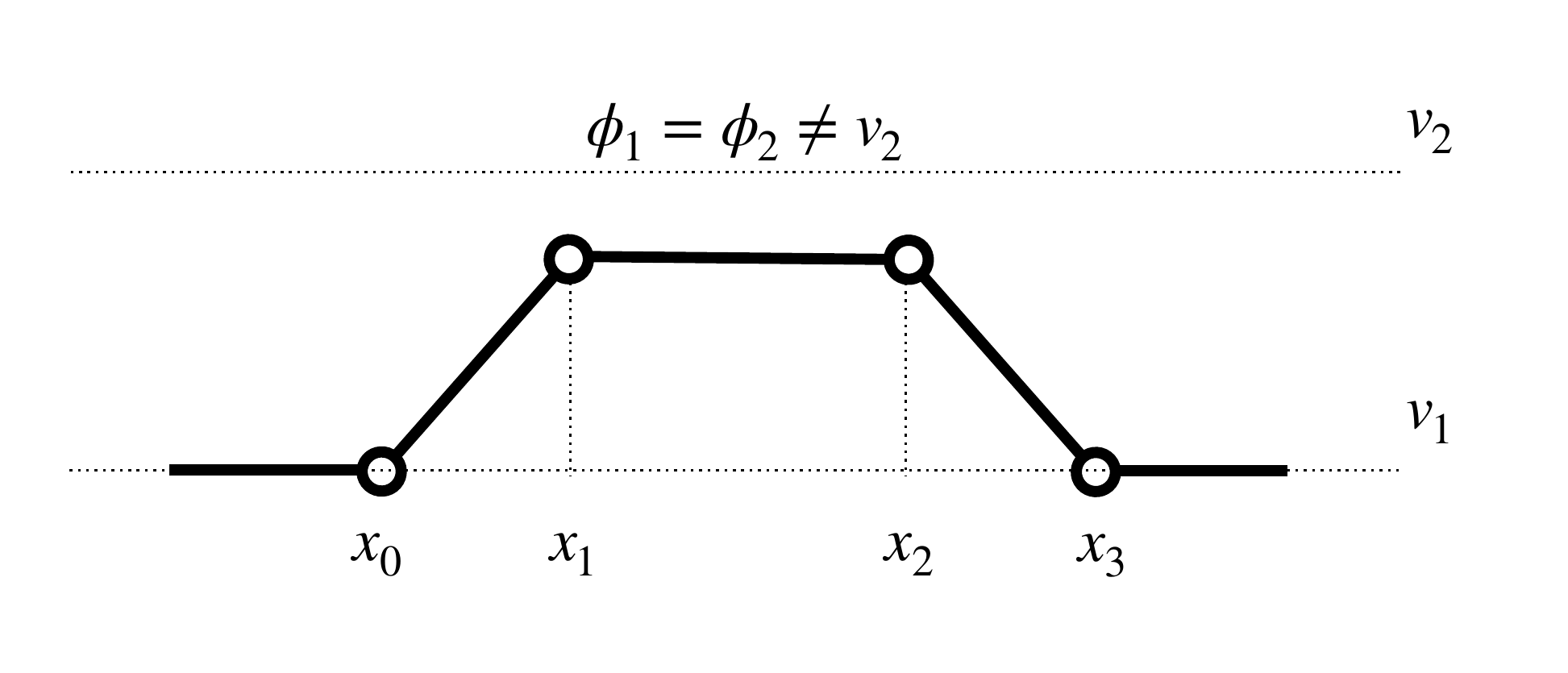}
\end{center}
{
    \caption{\small $N=3$ symmetric mech-field.}
    \label{fig:n3mech}
}
\end{figure}

One way around this obstacle is to investigate approximate mech-kinks. 
As an example, we can consider a symmetric $N=3$ mech-field depicted in Fig.~\ref{fig:n3mech} with a middle segment not exactly in a vacuum but arbitrarily close to it. 

However, the excess energy in the central segment would manifest as a constant attractive force between the (approximate) mech-kinks. Importantly, this force would be long-range and, clearly, an artefact of the selected configuration. In fact, we should not see the mech-field in Fig.~\ref{fig:n3mech} as describing well-separated mech-kinks -- the presence of a long-range force between them makes them manifestly not well-separated. Rather, we should say that Fig.~\ref{fig:n3mech} depicts a symmetric mech-oscillon and has (a priori) nothing to do with mech-$K\bar{K}$ dynamics.

Even for more elaborate mech-fields, long-range forces would remain present simply because piece-wise linear functions cannot rapidly approximate a constant (without actually being identical to it), unlike the exponentially decaying tails of kinks in the field theory. 
In this way, the investigation of `approximate' mech-$K\bar{K}$ scattering is irreparably contaminated by unphysical longe-range forces.

Fortunately, we can investigate scattering of mech-kinks directly and in a natural way, as we shall do in this section. 

\subsection{Rigid $K\bar K$-scattering}

Let us consider a superposition of a mech-kink and anti-mech-kink with widths $R_K$ separated by a flat middle segment of length $R$ in a vacuum $v_2$. Specifically, 
\begin{widetext}
{\small \begin{align}
\phi_M^{K\bar{K}} =\, & v_1 \theta(-x-R_K -R/2) + \Bigl(\theta(x+R_K+R/2)-\theta(x+R/2)\Bigr)\Bigl(\frac{v_2-v_1}{R_K}(x+R_K+R/2)+v_1\Bigr)
+v_2\Bigl(\theta(x+R/2)-\theta(x-R/2)\Bigr)\nonumber \\ \label{eq:mechkak}
& +\Bigl(\theta(x-R/2)-\theta(x-R_K-R/2)\Bigr)\Bigl(-\frac{v_2-v_1}{R_K}(x-R/2)+v_2\Bigr)
+v_1 \theta(x-R/2-R_K)\,.
\end{align}}
\end{widetext}
Note that $\phi_M^{K\bar{K}} = \phi_M^{K}(x+R/2)+\phi_{M}^{\bar{K}}(x-R/2)+v_1$, where $\phi_M^{K, (\bar K)}$ is an $N=1$ (anti-)mech-kink.

\begin{figure*}[htb!]
\begin{center}
\includegraphics[width=0.9\textwidth]{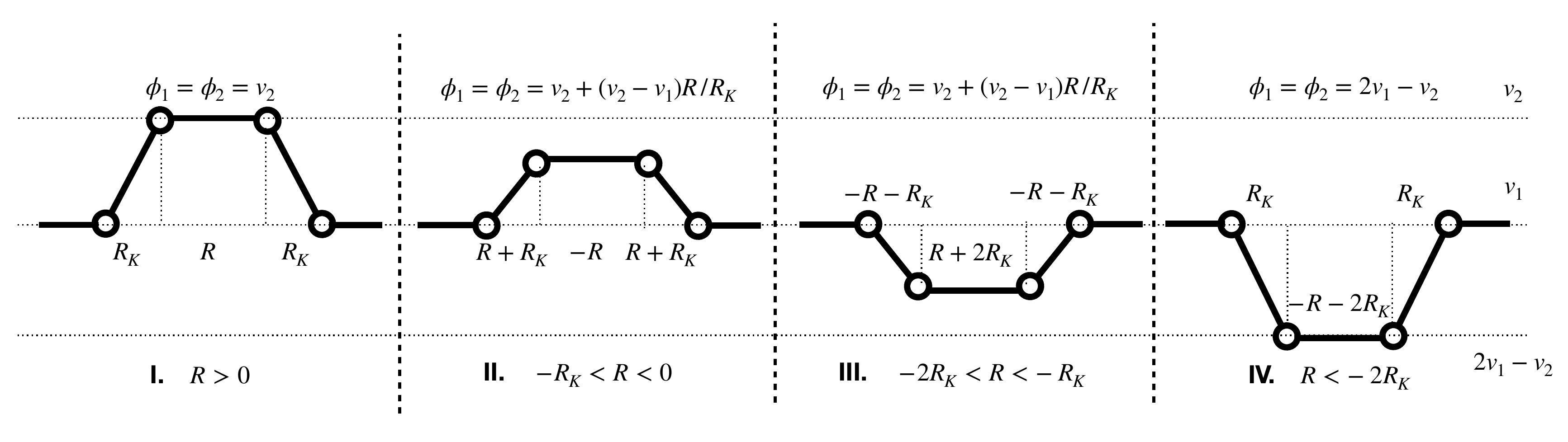}
\end{center}
{
    \caption{\small Various `stages' of $N=3$ mech-field depending on the value of $R$. If $R>0$, the configuration is a mech-kink and anti-kink interpolating vacua $v_1$ and $v_2$ separated by distance $R$. The width $R_K>0$ is assumed to be always positive. In the second and third stages, when $-2R_K < R <0$, the mech-field becomes a $N=3$ mech-oscillon, first above and then below the vacuum $v_1$. Its amplitude depends on $R$ in a designated way. In the final stage $R< -2R_K$, the mech-field can be arbitrarily wide, but since the middle segment does not lie in either vacua (assuming $2v_1-v_2$ is not another vacuum), it is energetically unfavored.}
    \label{fig:semirigid}
}
\end{figure*}

The key to analyze the dynamics of $\phi_M^{K\bar{K}}$ is to realize that 
there are four distinct orderings of joints (or \emph{stages}, see Fig~\ref{fig:semirigid}) depending on the value of $R$, each of which is governed by a different effective Lagrangian. 

Each stage follows from the previous one by continuation of $R$ to more negative values. In turn, $R$ has a different meaning in each stage. When $R>0$, it is the distance between mech-kink and anti-kink. When $-R_K < R <0$, it controls both the height and width of a symmetric $N=3$ mech-oscillon that is placed above vacuum $v_1$, while, when $-2R_K<R<-R_K$, $R$ serves a similar role for the mech-oscillon placed under $v_1$. Finally, when $R <-2R_K$, the mech-field $\phi_M^{K\bar{K}}$ resembles arbitrary separated anti-mech-kink and mech-kink that interpolates the values $v_2$ and $2v_1-v_2$. If the latter value is not a vacuum of the given model, there is a constant, attractive force and the whole configuration is energetically  disfavoured.

For simplicity, in this subsection we investigate \emph{rigid} mech-$K\bar K$ scattering, where only $R$ is dynamical while $R_K$ is fixed to an appropriate initial value.

For the first stage, i.e. assuming $R>0$, we have:
 \begin{align}
L_{\rm I}
= &\ \frac{(v_2-v_1)^2}{4R_K}\dot{R}^2-\frac{(v_2-v_1)^2}{R_K}-2 \kappa R_K\,,
\end{align}
where $\kappa$ is given either as in Eq.~\refer{eq:kappa} for non-BPS mechanization or as in Eq.~\refer{eq:bpskappa} in the BPS case. Up to an irrelevant constant, $L_{\rm I} $ describes a free particle (with position $R/2$) of mass $2(v_2-v_1)^2/R_K$. \footnote{If we replace $R_K$ with an appropriate static value this would be equivalent to twice the mass of $N=1$ mech-kink} In this stage, the mech-kinks indeed behaves as free particles.

In the second stage, i.e. for $-R_K < R <0$, we obtain
 \begin{align}
L_{\rm II} & =\frac{(v_2-v_1)^2}{4R_k^2} \bigl(R_K-R\bigr) \dot R^2- U_{\rm II}\,, 
\end{align}
where the potentials for both non-BPS and BPS mechanization read
{\small \begin{align}
U_{\rm II} = & \frac{2R_K}{v_2-v_1}\Bigl(\mathcal{V}\Bigl(v_2+\frac{v_2-v_1}{R_K}R\Bigr)-\mathcal{V}(v_1)\Bigr)
\nonumber \\ & -R V\Bigl(v_2+\frac{v_2-v_1}{R_K}R\Bigr) -\frac{(v_2-v_1)^2(R+R_K)}{R_K^2}\,, \\
U_{\rm II}^{\rm BPS} & =  \frac{R_K^2/(v_2-v_1)^2}{ (R_K+R)}\Bigl(\mathcal{W}\Bigl(v_2+\frac{v_2-v_1}{R_K}R\Bigr)-\mathcal{W}(v_1)\Bigr)^2\nonumber \\
-&\frac{R}{2}W^2\Bigl(v_2+\frac{v_2-v_1}{R_K}R\Bigr) -\frac{(v_2-v_1)^2(R+R_K)}{R_K^2}\,.
\end{align}}
 It is easy to check that $L_{\rm I} = L_{\rm II}$ at $R=0$, namely that the transition from first stage to second stage is \emph{continuous}. This is a direct consequence of continuity of the mech-field $\phi_M^{K\bar{K}}$ \refer{eq:mechkak}.

The Lagrangian $L_{\rm II}$ describes a rigid and symmetric $N=3$ mech-oscillon. Its solutions are periodic. Also notice, that the `metric'  $(v_2-v_1)^2 (R_K-R)/(2R_K^2)$ is regular at both transitions from the first to second stage ($R=0$) and from the second to third stage at $R= -R_K$. This is also true for the potential which goes to zero  at $R=-R_K$. Thus, there are no singularities that prevent us to further continue the variable $R$ below $-R_K$. 

The third stage also depicts a rigid $N=3$ mech-oscillon which is placed below the vacuum $v_1$:
 \begin{align}
L_{\rm III} & = \frac{(v_2-v_1)^2}{4R_k^2}\bigl(3R_K+R\bigr)\dot R^2 
  - U_{\rm III}\,,
\end{align}
where
{\small \begin{align}
U_{\rm III}  = &  -\frac{2R_K}{v_2-v_1}\Bigl(\mathcal{V}\Bigl(v_2+\frac{v_2-v_1}{R_K}R\Bigr)-\mathcal{V}(v_1)\Bigr)
\nonumber \\ +& (2R_K+R) V\Bigl(v_2+\frac{v_2-v_1}{R_K}R\Bigr)+\frac{(v_2-v_1)^2(R+R_K)}{R_K^2}\,, \\
U_{\rm III}^{\rm BPS} & = - \frac{R_K^2/(v_2-v_1)^2}{ (R_K+R)}\Bigl(\mathcal{W}\Bigl(v_2+\frac{v_2-v_1}{R_K}R\Bigr)-\mathcal{W}(v_1)\Bigr)^2\nonumber \\
+& \frac{2R_K+R}{2}W^2\Bigl(v_2+\frac{v_2-v_1}{R_K}R\Bigr)+\frac{(v_2-v_1)^2(R+R_K)}{R_K^2}\,.
\end{align}}
Again, it is easy to check that $L_{\rm II} = L_{\rm III}$ at $R=-R_K$.
Furthermore, both the kinetic term and the potential term remain well-defined at $R= -2R_K$, thus we may continue to the fourth stage:
 \begin{align}
L_{\rm IV} & =\frac{(v_2-v_1)^2}{4R_K}\dot{R}^2-\frac{(v_2-v_1)^2}{R_K} -2 \kappa_{\rm IV} R_K\,, 
\end{align}
where
{\small \begin{align}
\kappa_{\rm IV} = & \frac{\mathcal{V}(v_1)-\mathcal{V}(2v_1-v_2)}{v_2-v_1}-\Bigl(1+\frac{R}{2R_K}\Bigr)V(2v_1-v_2)\,, \\
\kappa_{\rm IV}^{BPS} & =  \frac{1}{2}\Bigl(\frac{\mathcal{W}(v_1)-\mathcal{W}(2v_1-v_2)}{v_2-v_1}\Bigr)^2\nonumber \\
&\phantom{=} -\frac{1}{2}\Bigl(1+\frac{R}{2R_K}\Bigr)W^2(2v_1-v_2)\,.
\end{align}}
The $L_{\rm IV}$  generically describes a particle in a linearly attractive potential, unless $2v_1-v_2$ is also a vacuum of the model.

Since all transitions from one stage to the next are continuous, we may collect the fixed-order Lagrangians into a single LOose Order Mechanical Lagrangian (LOOM) that captures the dynamics of rigid mech-$K\bar K$ collisions for the entire range $R\in (-\infty, \infty)$: 
\begin{align}
L_{\rm OOM}^{\rm rigid} = & \    L_{\rm I} \theta(R)+L_{\rm II}\theta(-R)\theta(R+R_K) \nonumber \\ &+
L_{\rm III}\theta(-R-R_K)\theta(R+2R_K)\nonumber \\ \label{eq:loomrigid}
&+L_{\rm IV}\theta(-R-2R_K)\,,
\end{align}
where $L_{\rm I, II, III, IV}$ are fixed-order Lagrangians given above. Note that $L_{\rm OOM}^{\rm rigid} $ is a continuous function of $R$. Furthermore, $L_{\rm OOM}^{\rm rigid} $ is nothing but a direct integration of the mech-field $\phi_M^{K\bar{K}}$ assuming $R_K = \mbox{const} >0$, i.e.
\begin{equation}
L_{\rm OOM}^{\rm rigid} = \int\limits_{-\infty}^{\infty}\diff x\, \mathcal{L}\bigl(\phi_M^{K\bar{K}}\bigr)\,,
\end{equation} 
\emph{allowing for all possible orderings of joints}.

\begin{figure}[htb!]
\begin{center}
\includegraphics[width=0.9\columnwidth]{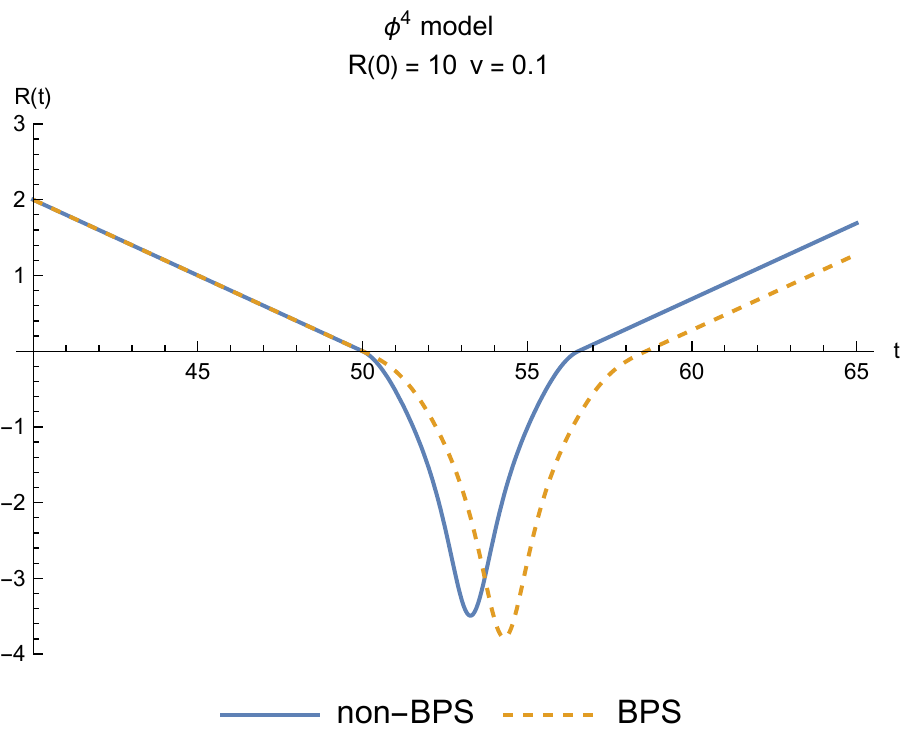}
\end{center}
{
    \caption{\small An example of a rigid mech-$K\bar{K}$ collision in $\phi^4$ model.}
    \label{fig:rigid4}
}
\end{figure}

\begin{figure}[htb!]
\begin{center}
\includegraphics[width=0.9\columnwidth]{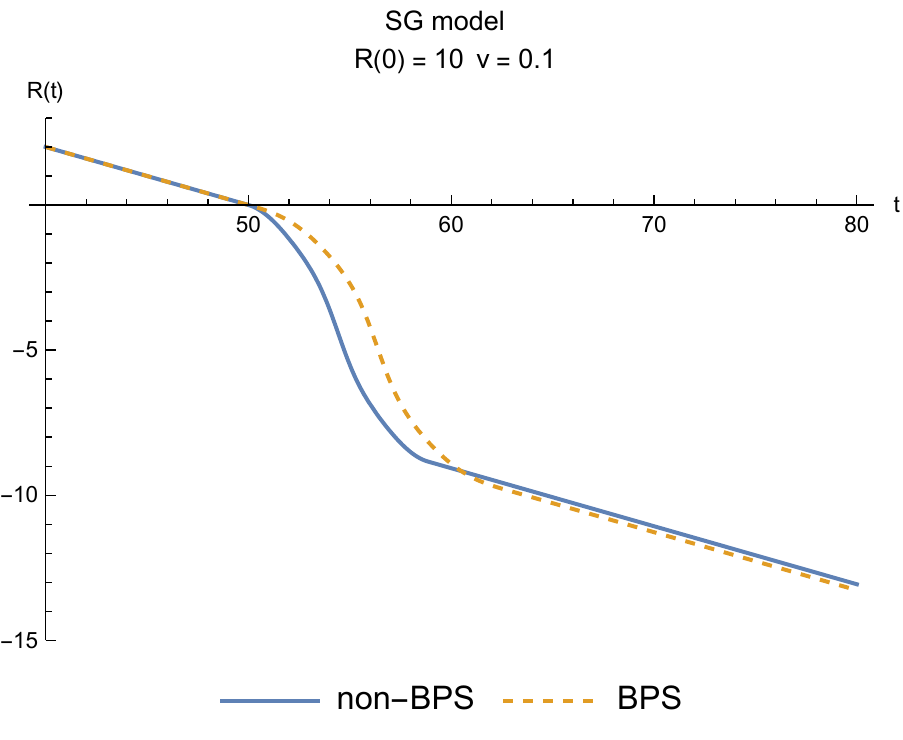}
\end{center}
{
    \caption{\small An example of a rigid mech-$K\bar{K}$ collision in SG model.}
    \label{fig:rigidsg}
}
\end{figure}

As there are no modes through which (rigid) mech-kinks can lose energy, they are either bound to reflect off each other (in case of, say, $\phi^4$ model) or go through each other (in case of, e.g., SG model) as illustrated on Figs.~\ref{fig:rigid4}-\ref{fig:rigidsg}.

Let us further point out that there is a direct field-theoretical analog of $L_{\rm Rigid}$ based on the ansatz
\begin{equation}\label{eq:contansatz}
\phi = \phi_K\bigl(x+R(t)/2\bigr)+\phi_{\bar K}\bigl(x-R(t)/2\bigr)+v_1\,,
\end{equation}
where $\phi_K$ is the single-soliton solution for the given model. 
The corresponding effective Lagrangian does the same job as $L_{\rm Rigid}$ of Eq.~\refer{eq:loomrigid}. However, there are conceptual differences. In field theory, one must be careful about placing the kinks sufficiently apart. This is especially true for the $\phi^8$ model where the kinks have long polynomial tails and naive superposition ansatz like \refer{eq:contansatz} is not suitable for numerical investigations \cite{Christov:2018wsa}. In contrast, the mech-$K\bar{K}$ superposition \emph{is} an exact solution of the equations of motion for any $R>0$. Thus the outcome of the collision is automatically independent on the initial separation. We can summarize this by saying that the short-range interactions of field theory are replaced  by \emph{contact} interactions in mechanization. 

\subsection{mech-$K\bar K$ scattering}

Let us now turn on $R_K(t)$. The mech-kinks now possess a Derrick mode that allows the resonant transfer of kinetic energy which manifests as \emph{bouncing} -- this is when mech-$K\bar K$ pair temporarily reemerges from the second stage into the first stage but does not have sufficient kinetic energy to fly to apart and, instead, plunge again bellow the $R=0$ line. Notice that this gives us an operational definition of the number of bounces -- it is the number of zeros of $R(t)$ divided by 2 minus one.\footnote{This is irrespective of what the outcome of the scattering is.}

With dynamical $R_K(t)$, the $N=3$ mech-oscillon in the second stage can decay, i.e. its width grows exponentially with time, while the height exponentially decreases (see \cite{Blaschke:2022fxp} for details). Hence, non-rigid mech-kinks display more complicated behavior with two primary outcomes: a well-separated mech-$K\bar{K}$ pair with excited Derrick modes or a state of decayed mech-oscillon.

The first stage is given as
 \begin{equation}
 L_{\rm I}^{K \bar{K}} = \frac{1}{2}g_{ab}^{\rm I}\dot X_a\dot X_b -\frac{(v_2-v_1)^2}{R_K}-2\kappa R_K\,,
 \end{equation}
 where $X_a = \{R,R_K\}$ and where the metric reads
 \begin{equation}
 g^{\rm I} = \frac{2}{3R_K}\begin{pmatrix} 3 & 3 \\ 
 3 & 4
 \end{pmatrix}\,.
 \end{equation}
 There are no singularities in either metric or potential along the $R$ coordinate.
 
 The second stage is described by 
 \begin{equation}
 L_{\rm II}^{K \bar{K}} = \frac{1}{2}g_{ab}^{\rm II}\dot X_a\dot X_b -U_{\rm II}\,,
 \end{equation}
 where $U_{\rm II}$ is the same as in previous subsection and where the metric reads
 \begin{equation}
g^{\rm II} = \frac{1}{3R_K^4}
\begin{pmatrix}
6 R_K^2 \left(R_K-R\right) & 6 R_K \left(R_K^2+R^2\right) \\
 6 R_K \left(R_K^2+R^2\right) & -4 \left(R^3-2 R_K^3\right) \\
\end{pmatrix}\,.
 \end{equation}
 From the formulae for determinant and Ricci scalar
{\small \begin{gather}
 \left|g^{\rm II}\right|= -\frac{4 \left(R_K+R\right) \left(R^2 R_K+5 R R_K^2-R_K^3+R^3\right)}{3 R_K^6}\,, \\
 \mathcal{R} = \frac{9 R_K^4 \left(R^2 R_K-2 R R_K^2-R_K^3+R^3\right)}{2 \left(R_K+R\right)^2 \left(R^2 R_K+5 R R_K^2-R_K^3+R^3\right)^2}
 \end{gather}}
we see that there is a physical singularity at the transition into the third stage, i.e. at $R=-R_K$. Unlike in the rigid case, we are now unable to continue to more negative values -- the mech-field cannot attain values below the vacuum $v_1$. This is an unphysical artefact that can be traced to the so-called null-vector problem -- the metric is degenerate at $R=-R_K$.
 
 Hence, the full effective Lagrangian for mech-$K\bar{K}$ scattering has only two stages:
 \begin{equation}
 L_{\rm OOM}^{K\bar{K}} = L_{\rm I} \theta(R) + L_{\rm II}\theta(-R)\,.
 \end{equation}
 
A CCM that has similar characteristics as $L_{\rm OOM}^{K\bar{K}}$ is given by the ansatz
\begin{equation}
\phi = \phi_K\Bigl(b(t)(x+a(t))\Bigr)+\phi_{\bar{K}}\Bigl(b(t)(x-a(t))\Bigr)+v_1\,.
\end{equation}
The corresponding effective Lagrangian is described in \cite{Adam:2021gat} and suffers from the same null vector (or flatness) problem, namely that the metric has a singularity at $a =0$. There are also known remedies for this malady, either by  choosing different moduli space with a massive modes supplanted into the above ansatz with judiciously chosen amplitude moduli (see \cite{Manton:2021ipk}) or to include the Derrick modes in a perturbative fashion as done in \cite{Adam:2021gat}.

The flatness problem present in $L_{\rm OOM}^{K\bar{K}}$ is a consequence of too simple a mech-field. Indeed, from the formula \refer{eq:mechkak} we see that at $R=-R_K$ the mech-field becomes exact vacuum everywhere, i.e. $\phi_M^{K\bar{K}} = v_1$. Moreover, at this point $\partial_R \phi_M^{K\bar{K}} = -\partial_{R_K}\phi_M^{K\bar{K}}$. 

The null-vector problem, however, should all but disappear for higher $N$ mech-fields, where there are more degrees of freedom and it is easy to avoid $\phi_M = v_1$ for all values of $R$. We plan to investigate higher $N$ mech-$K\bar{K}$ collisions in the future.

\begin{figure*}[htb!]
\begin{center}
\includegraphics[width=0.99\textwidth]{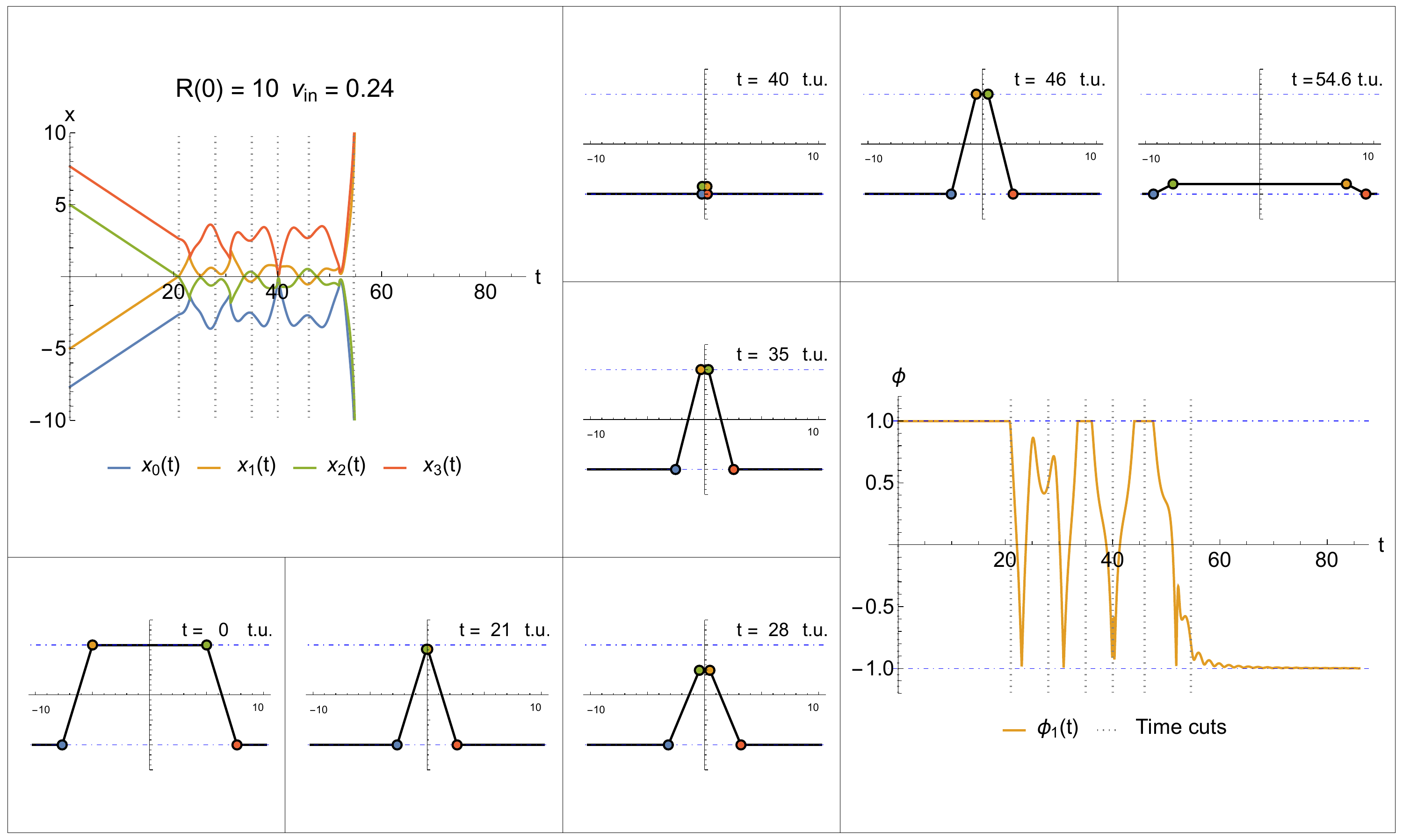}
\end{center}
{
    \caption{\small  Collision of mech-$K\bar{K}$ pair in $\phi^4$ model (non-BPS). Here we see a formation of a short-duration mech-oscillon. There are two bounces (around $t = 35$ and $t = 46$) after which the mech-oscillon rapidly decays into $-1$ vacuum.}
    \label{fig:gall1}
}
\end{figure*}

\begin{figure*}[htb!]
\begin{center}
\includegraphics[width=0.99\textwidth]{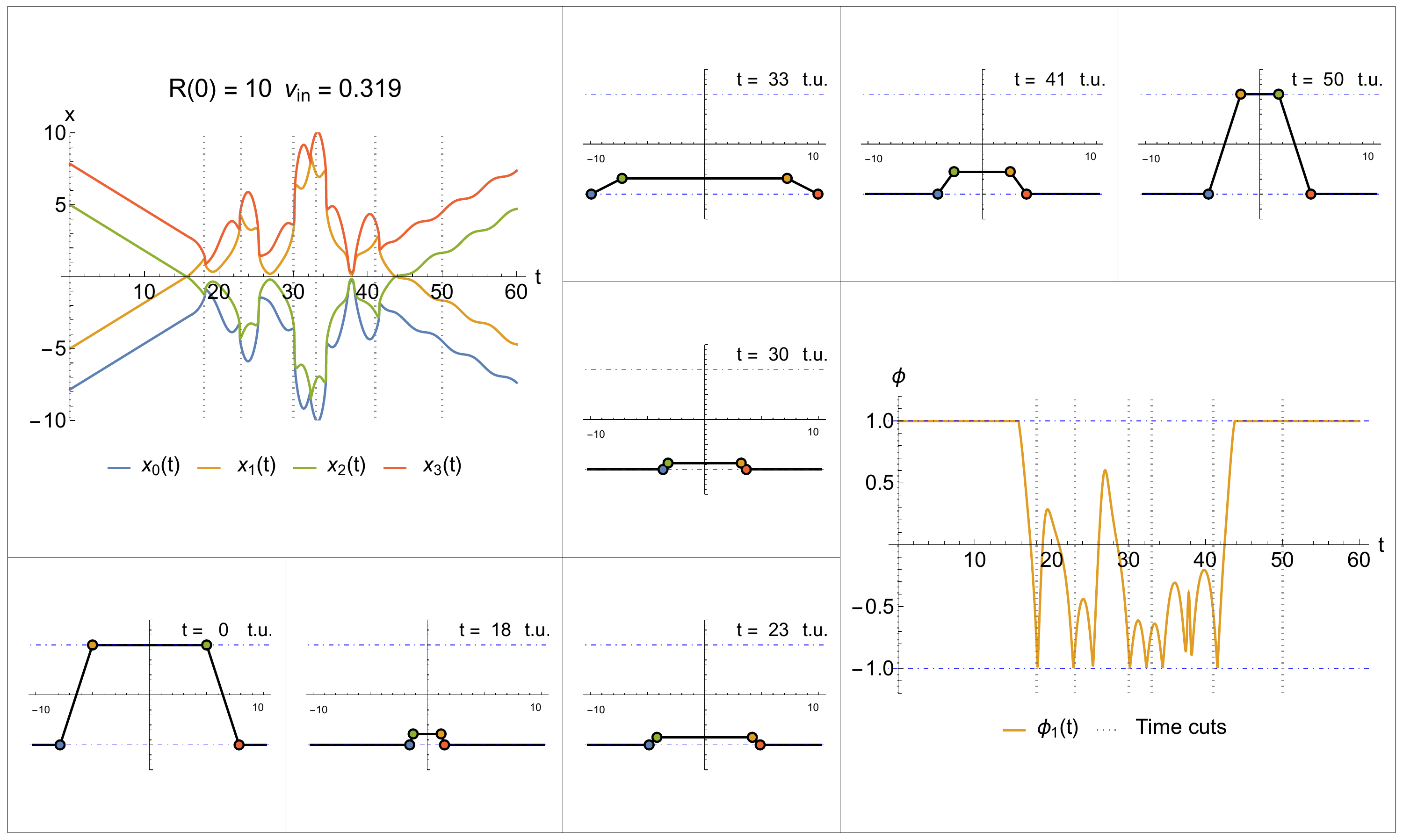}
\end{center}
{
    \caption{\small Collision of mech-$K\bar{K}$ in $\phi^4$ model (BPS). Here, the mech-oscillon is formed for roughly 25 t.u. and disintegrates into a excited mech-$K\bar{K}$ pair.}
    \label{fig:gall2}
}
\end{figure*}

\begin{figure*}[htb!]
\begin{center}
\includegraphics[width=0.99\textwidth]{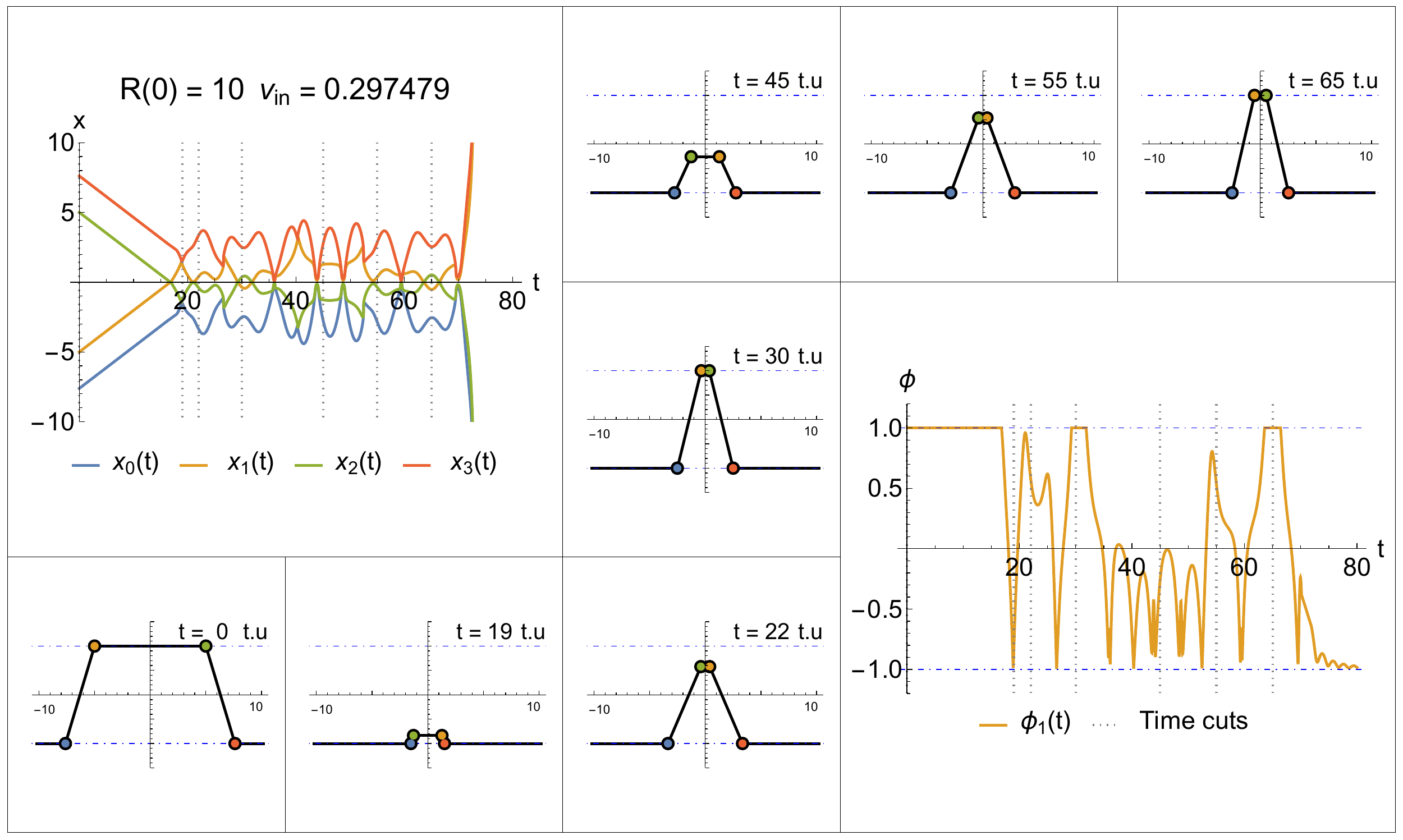}
\end{center}
{
    \caption{\small Collision of mech-$K\bar{K}$ in $\phi^4$ model (non-BPS). Here, we see two bounces (around $t = 22$ and $t=65$) after which the pair fly apart to infinity.}
    \label{fig:gall3}
}
\end{figure*}

We display typical mech-$K\bar{K}$ scattering on Figs.~\ref{fig:gall1}-\ref{fig:gall3}.

\begin{figure*}[htb!]
\begin{center}
\includegraphics[width=0.9\textwidth]{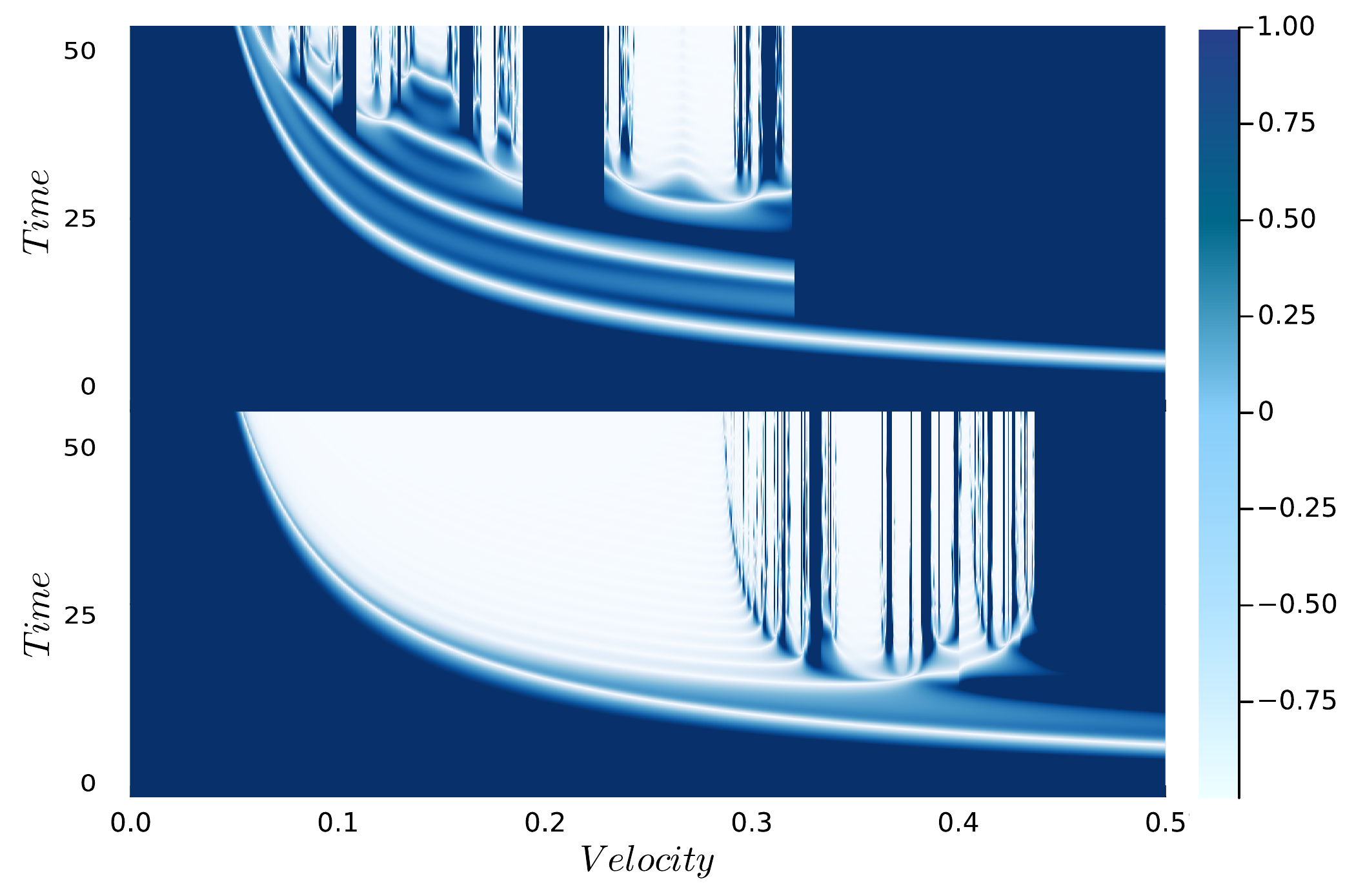}
\end{center}
{
    \caption{\small Dependence of the center value $\phi_M(x=0)$ on time for a range of initial velocities in the $\phi^4$ model. Upper part: non-BPS, lower part: BPS.}
    \label{fig:ten}
}
\end{figure*}

In Fig.~\ref{fig:ten} we display time-dependence of the center value of the mech-field $\phi_M(x=0)$ for a range of initial velocities in the $\phi^4$ model. 
The dark blue color represents $+1$ vacuum, while the white color stands for $-1$ vacuum. A bouncing occurs for such velocities when the value of $\phi_M(x=0)$ eventually returns to $+1$ value -- these are the dark blue columns indicating that mech-$K\bar{K}$ pair has been reformed after the initial collision. On the other hand, the white columns correspond to situations where the mech-oscillon decayed to -1 vacuum. 
The reader should compare Fig.~\ref{fig:ten} with Fig.~\ref{fig:themap}. 

Let us make several comments.

First, there exists a critical velocity above which the mech-$K\bar{K}$ pair scatter elastically. The corresponding value for $K\bar{K}$ collisions in $\phi^4$ theory is $v_{\rm crit} = 0.2598$ \cite{Adam:2021gat}, while for non-BSP $L_{\rm OOM}^{K\bar{K}}$ it is around 0.32 (upper half of Fig.~\ref{fig:ten}) and around 0.44 for BPS case (lower-half  of Fig.~\ref{fig:ten}). It is curious that BPS-ness makes the match with the field theory in this regard worse.
  
Second, there also exists a minimal velocity, below which bouncing does not occur. This is especially visible for BPS mechanization, where $v_{\rm min} \approx 0.28$ and below which there is a clear change in character of the collisions.  For non-BPS version (the upper part of Fig.~\ref{fig:ten}), it is much harder to ascertain the value of $v_{\rm min}$ without careful search. Certainly, there does not seem to be a qualitative change of the mech-$K\bar{K}$ scattering like in the BPS case. In this regard, the BPS case is more similar to the field theory.

The third aspect -- and perhaps the most striking one -- is the absence of smooth edges between bouncing windows and bion chimneys in both non-BPS and BPS cases. In field theory, bouncing windows begin and end at practically vanishing values of outgoing velocities so that the transitions to bion chimneys are continuous. In our case, however, bouncing windows start and end abruptly at finite velocities. Currently, we cannot give a reason why this is so nor what is the significance of this phenomenon. We suspect it being an clue that could lead to further conceptual improvements of the mechanization and we plan to investigate this aspects of mech-$K\bar{K}$ scattering in a future publication.

\subsection{LOOM}

We can generalize the concept of LOOM as follows. Given a mech-field, $\phi_M$, with joints placed at positions $\{x_0\,, \ldots\,, x_N\}$, the effective Lagrangian describing its dynamics is given by
 \begin{equation}\label{eq:loom}
L_{\rm OOM} = \sum\limits_{\sigma \in P_{N+1}} L_{\rm M}\bigl(\{x_{\sigma}, \phi_{\sigma}^{\rm true}\}\bigr)\prod\limits_{a=0}^{N-1}\theta\bigl(x_{\sigma(a+1)}-x_{\sigma(a)}\bigr)\,,
\end{equation}
where $P_{N+1}$ is the set of all permutations of $N+1$ indices and where $L_{M}(\{x,\phi\})$ is a fixed order effective Lagrangian given by the formula \refer{eq:efflag}. Notice that the (true) heights of the joints $\phi^{\rm true}$ depend on the given permutation of joints (e.g. in  Fig.~\ref{fig:semirigid} we see that the heights of joints have different values in each stage). Depending on which $L_M(\{x,\phi\})$ is taken, either non-BPS or BPS, the resulting LOOM would describe either non-BPS or BPS dynamics.

Let us stress that $L_{\rm OOM} $ is nothing else than a result of direct integration of the Lagrangian density, i.e.
\begin{equation}
L_{\rm OOM}  = \int\limits_{-\infty}^{\infty}\diff x\, \mathcal{L}\bigl(\phi_M\bigr)\,,
\end{equation}
allowing$\{x_0\,, \ldots\,, x_N\}$ to take all possible values. Consequently, the chain of Heaviside functions in the formula \refer{eq:loom} is there to enforce that the appropriate fixed-order Lagrangian is switched on.

Lastly, let us comment that a generic mech-field for which all $2N$ degrees of freedom are dynamical is not going to be directly extendable beyond canonical orderings of joints. This is due to the presence of singularities in the moduli space that we discussed in Sec.~\ref{sec:II} (see, e.g. \refer{eq:det2}). Generically, $\phi_a$'s need to be constrained in an appropriate way to ensure the continuity of the mech-field at each contact $\Delta x_a =0$.

\section{Summary and outlook}
\label{sec:V}

In this paper, we have introduced two conceptual advancements of mechanization compared with \cite{Blaschke:2022fxp}. 

First is the BPS mechanization that replicates the BPS nature of kinks in field theory. Unlike the non-BPS mech-kinks with static energies $m_K$ approaching the field theoretical mass $M_K$ only very slowly, as evident from Fig.~\ref{fig:four}, the BPS mech-kinks saturates the same bound $m_K = M_K$ for all $N$. There is also a major difference in the number of zero modes. Non-BPS mech-kinks have $2N-1$ massive normal modes and only one zero mode associated with translational symmetry. On the other hand, BPS mech-kinks have $N$ massive modes and $N-1$ additional zero modes that stem from the arbitrariness of heights of joints $\phi_a$'s. These extra zero modes make the $N>1$ BPS mech-kinks dynamically unstable. We observed that they are very prone to joint ejections -- a boundary joint flying off to infinity, effectively reducing the number of degrees of freedom. 
The joint ejections happen also for non-BPS mech-kinks, but in that case, there is an energy barrier to be overcome. This occurs almost spontaneously, as there is no energy barrier that needs to be overcome, unlike for the non-BPS mech-kinks. We are thus led to the conclusion that only the $N=1$ BPS mech-kink is a dynamically stable solution. 

We have also illustrated the convergence (or lack thereof) of the distribution of normal modes to the spectrum for kinks in Figs.~\ref{fig:five}-\ref{fig:convspec}. Especially Fig.~\ref{fig:convspec} testifies that  normal modes of BPS mech-kinks are unwilling to approximate the shape mode of $\phi^4$ kink even for very high $N$. This could be an important clue. Coupled with the observation of dynamical instability of $N>1$ BPS mech-kinks, we may conclude that there is a conceptual difference between kinks and mech-kinks that persists for any $N$. One difference that indeed persists is the compactness of the mech-field. Thus, it may be not enough just to increase $N$ to reach a quantitative match with field theory, but some new ingredient might be needed. We plan to pursue this clue in the future.

The compactness of mech-fields is also the reason for introducing the second conceptual advancement. In the pursuit of direct mech-$K\bar{K}$ scattering, we have shown that short-range interactions in field theory can be replaced by contact interactions by allowing joints to pass through each other. This can be achieved without encountering singularities, that are present in a general mech-field, by working with constrained mech-fields. Fortunately, a mech-$K\bar{K}$ pair separated by a flat segment of length $R>0$ is such a mech-field whose moduli space is regular at $R=0$ and we may continue the dynamics for negative values $R<0$ via switching to a (constrained) mech-oscillon as described in Sec.~\ref{sec:IV}. This is embodied in the notion of the LOose Order Mechanization, or LOOM.  

First, we have presented a LOOM for rigid mech-$K\bar{K}$ scattering, where only the separation $R$ is dynamic. There, the LOOM involves four stages (see Fig.~\ref{fig:semirigid}) that cover the full range of the moduli $R\in (-\infty, \infty)$. When we turned on the time dependence of widths, $R_K(t)$, we encountered a singularity at the transition from the second to the third stage preventing the mech-field to go below the vacuum. The same problem, known as the null-vector problem, appears also in naive CCMs and can be overcome by including more moduli \cite{Manton:2021ipk}. In our approach too, we expect the null-vector problem to go away when we increase the number of joints. We plan to provide details in a subsequent publication.

It is surprising that both BPS property and the necessity of LOOM point to the same conclusion: a generic mech-field has too much degrees of freedom and it needs to be constrained. This is perhaps not surprising, given the ad-hoc nature its construction. In particular, during the course of dynamical evolution, the $N>1$ BPS mech-kinks easily sheds degrees of freedom via joint ejections. This, in restricted capacity, is true also for non-BPS mech-kinks and it has been observed in high-$N$ mech-oscillons as well \cite{Blaschke:2022fxp}. The construction of LOOM -- which consists in sewing together effective Lagrangian for different orderings of the joints -- would not work at all if we allow all moduli of a generic mech-field to be dynamic. 

As a closing remark, however, let us also point out a case where the problem could be the exact opposite. Indeed, if we consider a free field theory with $V=0$, a general solution is described by superposition of arbitrary shapes moving with the speed of light either to the left or the right: $\phi_{\rm free}(x,t) = f_{\rm L}(x-t)+f_{\rm R}(x+t)$. As a corollary, the solution to the Cauchy problem with a static initial shape, i.e. $\phi_{\rm free}(x,0)=f(x)$ and $\dot \phi_{\rm free}(x,0) = 0$, is described by an immediate disintegration into two copies of the same shape with half the amplitude: $\phi_{\rm free}(x,t) = f(x-t)/2+f(x+t)/2$.  

In the mechanized version of the free field theory, this does not happen. Starting with a symmetric $N=2$ mech-field -- a triangle of width $R$ and height $A$, the governing equations of motion
\begin{equation}
\ddot R=\frac{16}{R}-\frac{2 \dot A \dot R}{A}\,, \quad
\ddot A=A\frac{ \dot R^2}{R^2}-\frac{20 A}{R^2}\,,
\end{equation}
can be solved exactly as 
\begin{equation}
R = R_0 +4 t\,, \quad A = A_0\sqrt{1+4t/R_0}\,,
\end{equation}
which depicts an ever-expanding triangle with a constant static mass $A^2/R = A_0^2/R_0$. Clearly, such a solution is unphysical, not just because the joints fly apart with twice the speed of light. The correct solution should be the same as in field theory -- that the triangle disintegrate to two similar triangles with half the initial height flying apart with the speed of light. However, this would require an instantaneous transition from symmetric $N=2$ mech-field with 3 joints to a symmetric $N=5$ mech-field with 6 joints. In other words, there would be a (triple) bifurcation at $t=0$ where each joint turns into two. Also note that such a bifurcation would be a relativistic effect.

We believe that some strange characteristics of mech-kinks dynamics can be potentially explained by the current lack of `bifurcation' process. We plan to pursue this possibility in a subsequent paper.

\acknowledgments

We would like to thank T.~Roma\'nczukiewicz and A.~Wereszczy\'nski for many fruitful discussions and help with numerical code.
We also acknowledge the institutional support of the Research Centre for Theoretical Physics and Astrophysics, Institute of Physics, Silesian University in Opava.


\end{document}